\documentclass[traditabstract,printer]{aa}
\usepackage{txfonts}
\usepackage{graphicx}
\usepackage{subfigure}
\usepackage{color}
\usepackage{natbib}
\usepackage{xfrac}

\begin{document}
\title{Purveyors of fine halos: Re-assessing globular cluster contributions to the Milky Way halo build-up with SDSS-IV
} 
\author{
Andreas Koch\inst{1} 
  \and Eva K. Grebel\inst{1}
   \and Sarah L. Martell\inst{2,3}
  }
\authorrunning{A. Koch et al.}
\titlerunning{Globular Clusters and Stellar Halo Build-up}
\offprints{A. Koch;  \email{andreas.koch@uni-heidelberg.de}}
\institute{Astronomisches Rechen-Institut, Zentrum f\"ur Astronomie der Universit\"at Heidelberg, M\"onchhofstr.\ 12--14, 69120 Heidelberg, Germany
\and
School of Physics, University of New South Wales, Sydney, NSW 2052, Australia
\and
Center of Excellence for Astrophysics in Three Dimensions (ASTRO-3D), Australia}
\date{}
\abstract{
There is ample evidence in the Milky Way for globular cluster (GC) disruption.
Hence one may expect that also part of the Galactic halo field star
population may once have formed in GCs. We seek to 
quantify the fraction of halo stars donated by GCs by searching for stars 
that bear the unique chemical fingerprints typical for a subset of GC 
stars often dubbed as ``second-generation stars''. These are stars showing
light element abundance anomalies such as a pronounced CN-band strength 
accompanied by weak CH-bands. 
Based on this indicator, past studies have placed the fraction of halo
stars with a GC origin between a few to up to 50\%.
Using low-resolution spectra from the most recent data release (DR14) of 
the latest extension of the Sloan Digital Sky Survey (SDSS-IV), 
we were able to identify  { 118 metal-poor ($-1.8 \le $ [Fe/H] $ \le -1.3$)} CN-strong stars in a sample of 4470 halo 
giant stars out to $\sim$50 kpc.  This increases the number of known 
halo stars with GC-like light-element abundances by a factor of { two}  
and results in an observed fraction of these stars of { 2.6$\pm$0.2\%.} 
Using an updated formalism to account for the fraction of stars lost 
early on in the GCs' evolution we thus estimate the fraction of the 
Galactic halo that stems from disrupted clusters to be 
very low, at 11$\pm$1\%. This number would represent the case that stars lost from GCs were entirely from 
the first generation and is thus merely  {an upper limit.}
Our conclusions are sensitive to our assumptions of the mass lost early on 
from the first generation formed in the GCs, the ratio of 
first-to-second generation stars, and other GC parameters. 
We carefully test the influence of varying these parameters on the 
final result and find that, under realistic scenarios, the above fraction 
depends  on the main assumptions 
{at less than 10 percentage points.}
We further recover a { flat trend in this fraction with 
Galactocentric radius, with a marginal indication of a rise beyond 30 kpc} that 
could reflect the ex-situ origin of the outer halo as is also seen in 
other stellar tracers.
}
\maketitle 
%
%
%%%%%%%%%%%%%%%%%%%%%%%%%%%%%%%%%%%%%%%%%%%%%%%%%%%%%%%%%%%%%%%%%
%
%
\section{Introduction}

In the current picture of hierarchical structure formation, major
parts of galactic halos stem from the accretion of small, presumably
dark-matter dominated, subgalactic units akin to the progenitors of
today's dwarf satellites 
\citep{SearleZinn1978,Dekel1986,Bullock2005}. This channel of halo
formation is indeed seen in action via observations of satellite
disruption in and outside our own Milky Way (MW; e.g.,
\citealt{Ibata1994,Belokurov2006,McConnachie2009,Koch2012HCC,Ludwig2012,Shipp2018,Morales2018}, to name a few). 

Moreover, globular clusters (GCs) are progressively gaining
consideration as building blocks of yet unknown fractions of the
Galactic stellar halo, which is also bolstered by their observed
disruption in the tidal field of the MW via tidal tails in the
Galactic halo
\citep[e.g.,][]{Odenkirchen2001,Lee2004,Lauchner2006,Chun2010,Jordi2010,NiedersteOstholt2010,Sollima2011,Myeong2017,Navarrete2017,CarballoBello2018,Kuzma2018},
the detection of extratidal stars with a likely GC origin based on,
e.g., photometric filtering techniques or stellar abundances and
velocities
\citep[e.g.,][]{Jordi2010,Kunder2014Tidal,Anguiano2015,Anguiano2016,Navin2016,Simpson2017NGC1851,Kunder2018},
or the discovery of thin streams with possible GC progenitors
\citep[e.g.,][]{Grillmair2006,Grillmair2009,Bonaca2012,Bernard2014}. 

{GCs stand out in that
they show significant light chemical element variations over a broad mass range of cluster masses ,}  
as was 
first noted in the 1970s.  Prominent manifestations of these are
significant dispersions in Na and O
\citep[e.g.,][]{Osborn1971,Cohen1978,Carretta2009NaO}, and
bimodalities in their stellar CN abundances
\citep[e.g.,][]{Popper1947,Harding1962,Norris1979,Kraft1982,Harbeck2003,Kayser2008,Smolinski2011}.
Furthermore, a clear correlation between Mg and Al is seen, stemming
from the same proton-burning channels \citep{Gratton2001}. 
{All of these lead to clear (anti-)correlations rather than generic abundance
spreads alone, despite not all GCs showing Mg spreads, as recently reviewed by
\citet{Pancino2017}.
 To higher order, these findings} are complemented by later
burning stages in the first stellar generation causing Mg, Si, and Zn
to anti-correlate as well \citep[e.g.,][]{Hanke2017}. 
Such give-away abundance variations have to date been traced down to
the low-mass GC regime \citep{Bragaglia2017,Simpson2017ESO}, and were also found  in old GCs
in external galaxies
\citep[e.g.,][]{Mucciarelli2009,Mateluna2012,Colucci2014,Larsen2018},
as well as in massive, intermediate-age clusters in the Magellanic
Clouds \citep[e.g.,]{Hollyhead2017,Hollyhead2018}
{ and down to cluster ages of $\sim$2 Gyr \citep{Martocchia2018}.} 

It is nowadays accepted that these variations go alongside with
multiple stellar populations in GCs, with colour-magnitude
diagrams uniquely supporting the presence of two or more populations, 
possibly separated in age by several $\sim 100$ Myr\footnote{ 
For ancient GCs (ages$>$10 Gyr), an age difference of $\sim$200 Myr is possible, limited by the accuracy in age dating, whereas
for the $\sim$2 Gyr old cluster NGC 1978, \citet{Martocchia2018NGC1978} found  that 
the two populations are coeval to within an upper limit of 20 Myr.},
 and distinct in other
tracers such as variations in He content.  However, the origin of these populations
is as yet unclear as no model can currently explain all of the
observational constraints \citep[e.g.,][]{Bastian2018}.  In
particular, it remains unresolved whether GCs did indeed experience
two or more episodes of star formation resulting in the observed
abundance anomalies.  Nonetheless, we will follow the commonly adopted
terminology of calling stars that are enhanced in O and C and that
show lower Na and N abundances (in other words, ``normal'' stars by
metal-poor GC standards) ``first generation'' stars, and those with
elevated Na and N levels ``second generation'' stars.  We do so for
the sake of convenience and without implying that we favour a
particular formation scenario such as repeated star formation. 
Either way, the ratio of stars in the two main populations in each
cluster lies at $\sim 50$\% each, with a possible trend toward a higher
fraction of the first-generation stars with decreasing cluster mass
(\citealt{Milone2017}; cf.\ \citealt{BastianLardo2015}).

{Ad-hoc models of GC evolution that are designed to explain multiple populations via multiple epochs
of star formation} \citep[e.g.,][]{DErcole2008,Bastian2013}
predict that, during an initial phase of cluster dissolution, a large
fraction of up to 90\% of the first generation of cluster stars is
lost from a GC and released into the halo.
 Since these first generation stars, however,
bear the chemical imprint of {a standard, early chemical evolution}, 
they are indistinguishable from any given halo field star.
More generally, several studies suggest GCs have, on average, lost a
considerable fraction of their stars since they formed, e.g., around
two thirds according to \citet{Kruijssen2015} or about 75--80\% according
to \citet{Baumgardt2017} and \citet{Baumgardt2019}.  

Since first-generation stars that end up in the halo are chemically
indistinguishable from ``normal'' halo stars, the second-generation
stars are of particular interest since they can be traced through
their peculiar chemistry.  They are enhanced in N and Na and, in turn,
depleted in C and O, a pattern that is qualitatively consistent with
proton-capture reactions in a generation of polluters such as
intermediate-mass asymptotic giant branch (AGB) stars
\citep{DErcole2008}, fast-rotating massive stars
\citep{Decressin2007}, massive binaries \citep{deMink2009}, or super-massive stars
\citep{Gieles2018}. 

In order to assess the actual fraction of the MW halo that is made up
by former GC stars, we can use the empirical fact that the
aforementioned light element variations are commonly found in GCs, but
not in young open clusters or dwarf galaxy field stars, and only
rarely in the halo field \citep{Pilachowski1996,Geisler2007}.  If any
stars with second-generation-like chemistry (classified by their molecular band strengths as CN-strong and CH-weak) 
exist in the halo, this
indicates that they most likely originated from disrupting GCs. 

Large sky surveys such as the Sloan Digital Sky Survey (SDSS) have
permitted statistical investigations of the fraction of the MW halo
contributed by disrupted GCs  based on the discovery of CN-strong and
CH-weak field halo stars
\citep[e.g.,][]{Martell2010,Martell2011,Carollo2013,Tang2019}, with estimates ranging from 17--50\%, depending on 
the assumptions of the ratio of first-to-second generation stars in GCs 
and the  statistics of GC dissolution. 
Likewise, \citet{Carretta2010} and \citet{Ramirez2012}
found  candidates with second-generation-like Na- and O-abundance ratios across the
metal-poor tails of all MW components, while \citet{Lind2015}
identified one potential GC escapee in the Gaia-ESO Survey
\citep[GES;][]{Gilmore2012} of a few hundred halo stars, based on its
second-generation-like Mg and  Al abundances. This was extended
towards a larger fraction of five out of  the studied seven metal-rich field stars with Mg-Al
anomalies using the SDSS-III APOGEE Survey
\citep{Majewski2017,Fernandez-Trincado2017}.  Similarly,
\citet{Martell2016} employed APOGEE to identify five halo giants
enriched in N and Al.  These more recent studies typically arrive at
estimates of just a few per cent of halo stars with GC cluster origin,
suggesting that GC disruption is a minor contributor to the halo field
population.
On the modeling front, \citet{Schaerer2011} estimate a contribution of
5--8\% of low-mass first generation stars to the halo field
population, and possibly as high as 20\% if also second-generation
stars are accounted for.

While the Galactic bulge is not the target of our current study, we
note that also in the Galactic bulge N-rich field stars were
discovered \citep{Schiavon2017}. \citet{Schiavon2017} argue that if
the N-rich stars are indeed former GC stars, then the mass in
destroyed GCs exceeds the mass in surviving GCs by a factor of eight.

In our work, we seek to  make use of the latest extensions to the SDSS
\citep{Blanton2017} in order to improve the number statistics of
CN-strong stars in the MW halo.  While previous studies detected a few
tens of such stars in spectroscopic samples of several thousand
(leading to a fraction of CN-strong field stars of $\sim 2$\%;
\citealt{Martell2011}), the latest generations of massive
spectroscopic surveys allow for a potential increase in the number of
disrupted GC stars by factors of several.  
This paper is organized as follows: In Sect.~2 we describe the data
set and steps taken to define the bona-fide halo sample. Our spectral
index measurements to determine CN- and CH- band strengths are laid
out in Sect.~3, and the method to discriminate CN-strong from
CN-normal stars is presented in Sect.~4. Next, Sect.~5 is dedicated to
the formalism to determine the fraction of the halo that likely
originated from dissolved GCs. Our results and their limitations are then discussed in
Sect.~6. 
\section{Data and sample selection}
The first statistical endeavour towards the fraction of halo stars
born in GCs \citep{Martell2010} was based on the seventh data release
(DR7) of the SDSS, which already included the first important add-on
of the Sloan Extension for Galactic Understanding and Exploration
(SEGUE) with additional spectra of 240,000 stars \citep{Yanny2009},
taken with a resolving power of $R$$\sim$2000 over a wavelength range
of 3800--9200 \AA.  The subsequent work of \citet{Martell2011} built upon
more numerous data from the next extension within SDSS-III, i.e.,
SEGUE-2 (with an additional 120,000 stellar spectra). After applying a number of 
stringent selection criteria, this yielded the
discovery of 16 further CN-strong stars, adding to the 49 such objects
out of $\sim 2000$ regular halo stars from the previous study. 

In our current study, we employed data from the latest phase of the
SDSS-IV \citep{Blanton2017}, as drawn from its DR14  catalog
\citep{Abolfathi2018}.  SDSS-IV's low-resolution spectroscopy
primarily targeted galaxies and quasars from the SDSS project
``extended Baryon Oscillation Spectroscopic Survey'' (eBOSS;
\citealt{Dawson2016}), which contains on the order of 9\% stellar
spectra. Moreover, DR14 contains the previously observed spectra from
the earlier phases of the SDSS.  As in the previous works, we relied
on stellar parameter and metallicity measurements provided in the SDSS
catalogs, which are, in turn, based on the automated Segue Stellar
Parameter Pipeline \citep[SSPP;][]{Lee2008}.

In order to build a representative sample of halo red giants, we apply
the same selection criteria as \citet{Martell2010} 
{ in our SQL query to the SDSS.
That is, we preselect only stars with [Fe/H]$\le -1$ to focus on halo objects. 
We further demand
that the error on metallicity is $\sigma$[Fe/H]$\le$0.5 and  that at
least three of the ten independent metallicity determinations within
the  SSPP were flagged as reliable.
Furthermore, only objects with surface gravities constrained to log\,$g\le3$, and
respective errors below  $\sigma$\,log\,$g\le0.5$ were queried.}  Finally, the mean
signal-to-noise ratio (S/N) was required to lie above 20 per pixel, and
we employed a colour cut towards cooler giants using $(g-r)_0 \ge
0.2$, leaving a starting sample of 
$\sim$15000 halo giants.
% (depicted as contours in Fig.~1).
%
%

Next, we required a S/N above 15 per pixel { in the region of
4000--4100~\AA, as lower S/N ratios would cloud our 
measurements of the respective spectral indices below}. 
 The ensuing subsample was further refined 
by dividing the entire sample into metallicity bins of 0.2 dex each.
{ Within each bin, we constructed fiducial lines in  
diagrams of log\,$g$ vs. colour (which transpired to yield the clearest separation) 
and pruned the sample from contaminating AGB, main sequence,
and turn-off stars using a smoothed 3$\sigma$-cut along those ridge
lines. This is illustrated for each metallicity bin in Fig.~1.}
\begin{figure}[htb] \centering
\includegraphics[width=1\hsize]{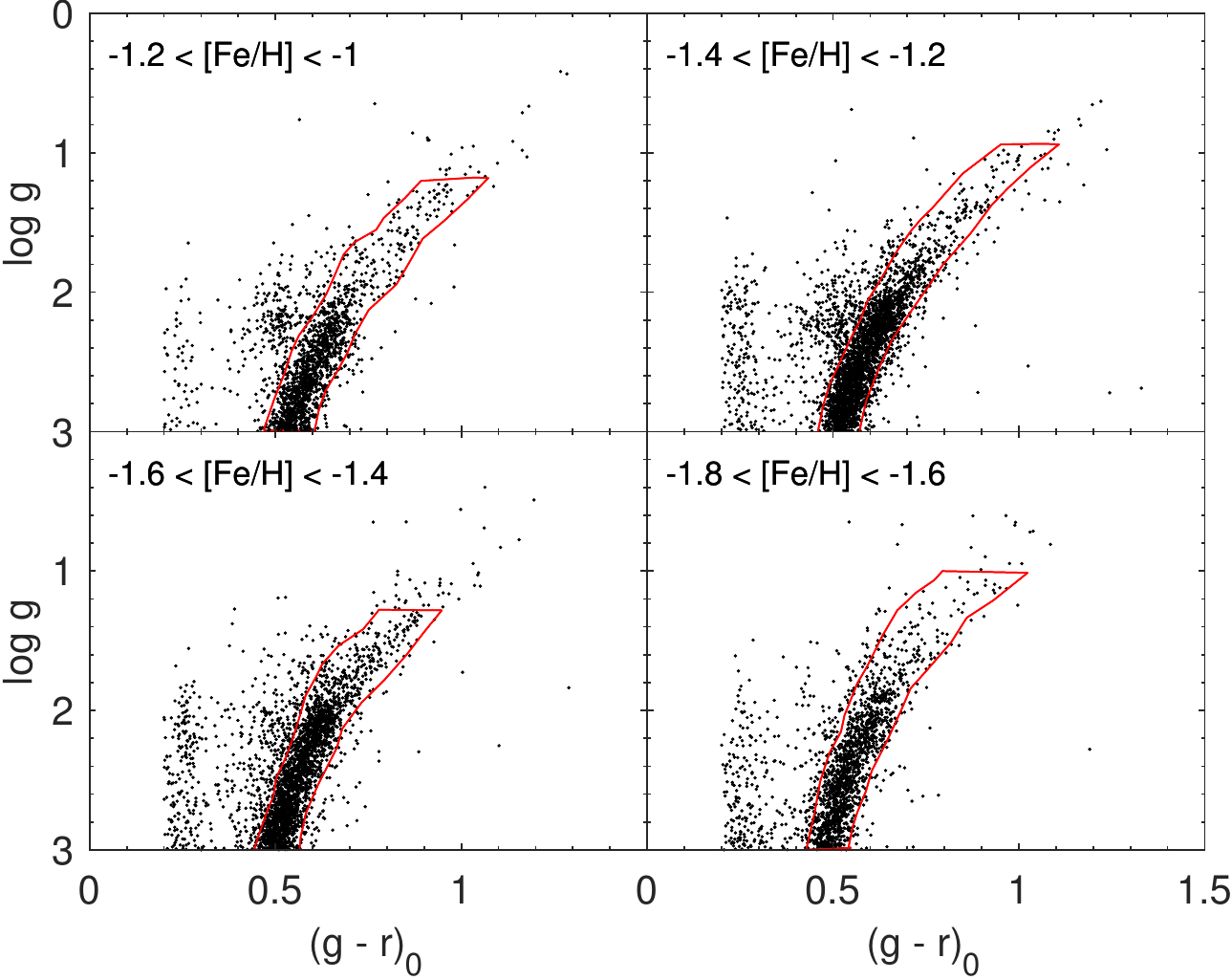} 
\caption{ Modified Kiel diagrams for several metallicity bins. The red 3$\sigma$-envelopes were used to 
select {\em bona-fide} halo giants.} 
\end{figure}

We also ascertained that our field star sample was uncontaminated by
present-day GC member stars, as they have not yet been released into
the surrounding halo.  To this end, we rejected stars that fall within
the tidal radii of any object from the catalog of \citet[][2010
version]{Harris1996}. In this way, we identified 43 giants located in
M3, M13, and M15.  
We were thus left with 6801 halo field candidates.

{ Finally, in the following, we explicitly ignored the metal-rich ([Fe/H]$> -1.3$ dex) tail of our distribution. 
This is to avoid any possible false detections of CN-strong stars that are not 
the progeny of actual GC dissolutions, as suggested by 
follow-up observations of a subsample of the CN-strong halo stars by \citet{Martell2011}, which do
not show any signatory Na-enhancement (Martell et al. in prep.).
While it is well established that metal-rich GCs do show CN/CH anti-correlations \citep{Harbeck2003,Kayser2008}
it is strictly not required that a former GC star would need to show both enhanced N and Na-enhancement for 
a first identification, in particular since, here, we are solely concerned with  low-resolution spectra 
without any means to constrain the Na content of our CN-strong stars. 
Nonetheless, imposing our additional metallicity cut will ascertain a {\em bona-fide} sample 
for unbiassed identification of the desired GC-like stars halo stars. 
Furthermore, metal-poor stars below $-1.8$ dex were removed from the sample as elaborated below in Sect.~2.1.
This leaves us with a final sample of 4649 stars with halo characteristics in the metallicity range of $-1.3$ to $-1.8$. 
Fig.~3 shows the full, queried initial and the selected halo subsample. 
}
\begin{figure}[htb] \centering
\includegraphics[width=0.78\hsize]{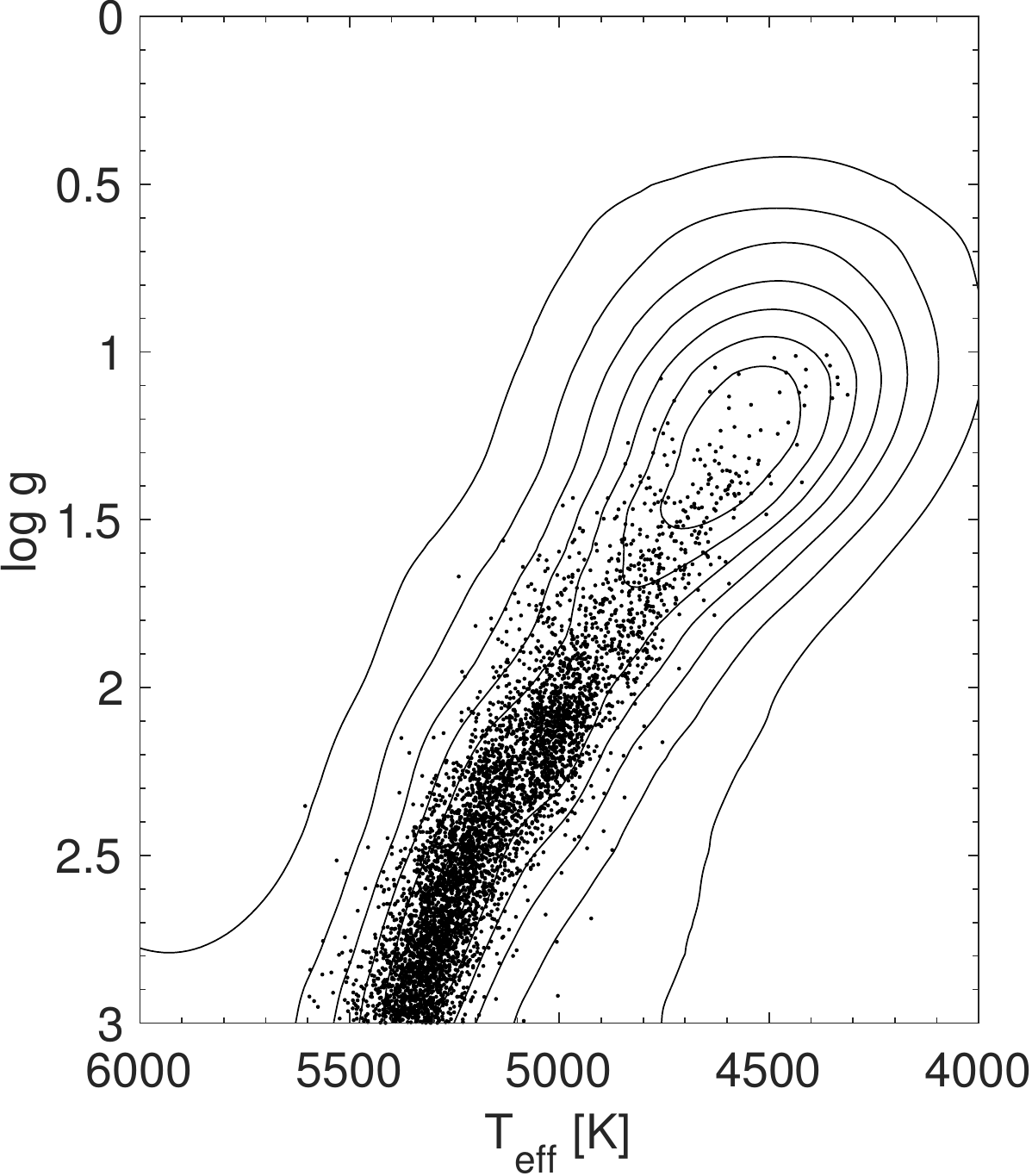} 
\caption{Kiel diagram of the full sample { (contours; satisfying the initial query restrictions in 
log\,$g$$<$3, [Fe/H]$<$$-1$, $(g-r)_0>0.2$, S/N$>$20, and respective cuts in the parameters' errors).  
and the subset of halo giants (dots; after removal of metallicity-dependent CMD features and further 
pruning the metal-rich and -poor tails above $-1.3$ and below $-$1.8 dex)} used for our statistical purposes.} 
\end{figure}
\subsection{Distances} 
In order to determine the stars' distances we first computed their
$r$-band absolute magnitudes, $M_r$, where we resorted to a set of
10-Gyr old Dartmouth isochrones \citep{Dotter2008} for a broad grid of
metallicities from $-1$ to $-2.5$ dex, using no $\alpha$-enhancement,
as these provide the best, global representation of metal-poor Galactic GCs 
 \citep[see also][]{Hendricks2014b}.  These isochrones were
interpolated to the observed, dereddened $(g-r)_0$ colour as provided
by the SDSS.  Since the isochrone spacing is getting smaller and less
distinguishable towards the lowest metallicities, and with the grid
limitation of the Dartmouth tracks to $\ge-$2.5 dex, we follow
\citet{Martell2011} in selecting only stars more metal-rich
than $-1.8$ dex, which mainly picks up halo giants broadly located
around the peak of the halo metallicity distribution function
\citep{Schoerck2009}. This is further beneficial, as the CN band
strength loses its sensitivity to N-abundance at low metallicities
\citep[e.g.,][] {Boberg2016M53}, leaving CN bimodalities largely
undetetectable from this tracer.
\section{Spectral indices}
At the low resolution of the SDSS spectra, measuring band indices in
the blue spectral region provides a powerful tool to identify
CN-strong and CN-weak stars.
In particular, the frequently used S(3839) index, first defined and
optimized by \citet{Norris1981} to measure the strength of the CN-band
at 3839~\AA, is ideal to trace  bimodal band strength distributions:
\begin{equation} 
S(3839)\,=\,-2.5\,\log \frac{\int_{3846}^{3883}
I_{\lambda}\,{\rm d}\lambda}{\int_{3883}^{3916} I_{\lambda}\,{\rm
d}\lambda}, 
\end{equation}
where $I_{\lambda}$ simply refers to the observed flux in the
spectrum. 

The run of this CN index versus { absolute magnitude (bottom panel), colour (middle), and metallicity
(top)} is shown in Fig.~2, displaying the trade-mark increase of
the CN-band strengths with colour, due to colder (thus redder) stars forming
progressively more (CN) molecules.  
\begin{figure}[htb] 
\centering
\includegraphics[width=1\hsize]{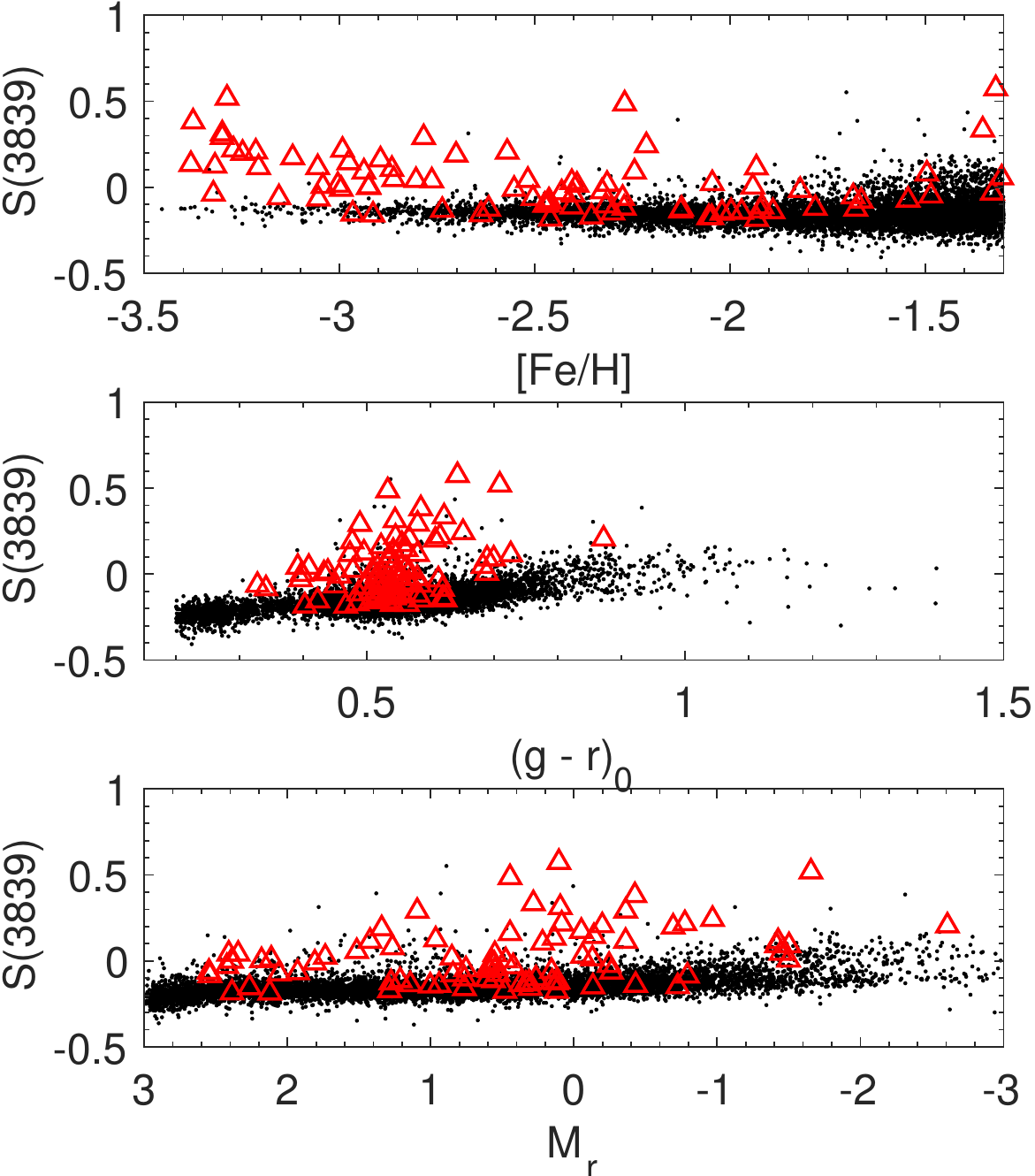} 
\caption{CN index in the halo giant candidates (black dots) according
to Eq.\ 1 versus { absolute magnitude (bottom panel, colour (middle) and metallicity (top)}. Carbon-
and CEMP-stars (red triangles) were rejected based on their $s(c_0)$
and $s(c_1)$ indices (Eqs. 2,3).} 
\end{figure}

Similarly, strong bands of carbonaceous molecules such as CH and CN
are seen in carbon stars, and, in the metal-poor regime, in
carbon-enhanced metal-poor (CEMP) stars
\citep{BeersChristlieb2005,CJHansen2016}. In this case, the strong
bands are rather a result of their strong carbon-overabundance and do
not reflect the light element variations that result from the early
nucleosynthesis in the GCs to be engulfed into the halo.  These
objects can be efficiently flagged using indices encompassing the CH-
(at 4350~\AA) and C$_2$-bands (at 4737~\AA) that are mainly sensitive
to carbon and only little affected by nitrogen features in the
neighbouring spectral regions. These indices read: 
\begin{eqnarray}
s(c_0)&\,=\,&-2.5\,\log \frac{\int_{4370}^{4400} I_{\lambda}\,{\rm d}\lambda}{\int_{4330}^{4335} I_{\lambda}\,{\rm d}\lambda\,+\,\int_{4440}^{4460} I_{\lambda}\,{\rm d}\lambda}\\
s(c_1)&\,=\,&-2.5\,\log \frac{\int_{4660}^{4742} I_{\lambda}\,{\rm d}\lambda}{\int_{4585}^{4620} I_{\lambda}\,{\rm d}\lambda\,+\,\int_{4742}^{4800} I_{\lambda}\,{\rm d}\lambda}
\end{eqnarray}
Hence, we discard another 179 stars from further analysis due to their
strong C-bands, employing the set of restrictions in $s(c_0)$,
$s(c_1)$, and [Fe/H] outlined in \citet{Martell2010}.  Separating
these by metallicity yields fractions of C-rich stars that are fully
compatible with those found by \citet{Martell2010}, moreover
displaying a metallicity distribution function that directly traces
that of such objects in the Galactic halo as found in other,
large-scale surveys \citep[e.g.,][]{Lucatello2006,Carollo2012}. In
particular, the largest fraction of C-stars is found at low
metallicities, where 73\% of our sample lie below $-2$ dex -- the
formal boundary of CEMP stars \citep{BeersChristlieb2005}. 

Finally, in order to cross-identify the ``CN-strong'' nature of the
stars of interest as also being ``CH-weak'',  we determine the
strength of the CH G-band at 4300~\AA~via the index definition of
\citet{Martell2008}:
\begin{equation}
S(CH)\,=\,-2.5\,\log \frac{\int_{4280}^{4320} I_{\lambda}\,{\rm d}\lambda}{\int_{4050}^{4100} I_{\lambda}\,{\rm d}\lambda\,+\,\int_{4330}^{4350} I_{\lambda}\,{\rm d}\lambda}
\end{equation}
Throughout our entire work, errors on all measured parameters were
determined in a Monte Carlo sense by varying the input quantities
($g_0$, $r_0$, [Fe/H]) by their respective uncertainty.
\section{Separation of CN-weak and -strong stars}
In order to identify CN-normal and CN-weak stars, several methods have
been devised in the literature. For instance, \citet{Gerber2018}
employed separation by spline interpolation in CN-band strength vs.\
absolute magnitude space. However, their data have higher S/N and
smaller errors on individual measurements, and the confirmed
membership of their sample stars with a known GC facilitated an
accurate $M_V$ determination.  Furthermore, \citet{Boberg2016NGC6791}
discuss an accurate treatment of the index-measurement errors, albeit
at super-Solar metallicities, which are beyond the range of our halo
sample. 

For our purpose, we first removed any temperature, { gravity}, and metallicity
trend in the S(3839) measurements via a straight-line fit to the
CN-weak sequence in the S(3839) vs.\ { magnitude space in Fig.~3 (bottom)}.  Computing
a vertical distance of each individual star from the best fit then
yielded a ``corrected'' index $\delta$S(3839). 
Since the slope of this parameter space mildly depends on metallicity,
flattening towards the metal-poor regime, we performed the calculation
of the  $\delta$S(3839) values in independent bins of metallicity,
each 0.1 dex wide (Fig.~4). 
\begin{figure}[htb]
\centering
\includegraphics[width=1\hsize]{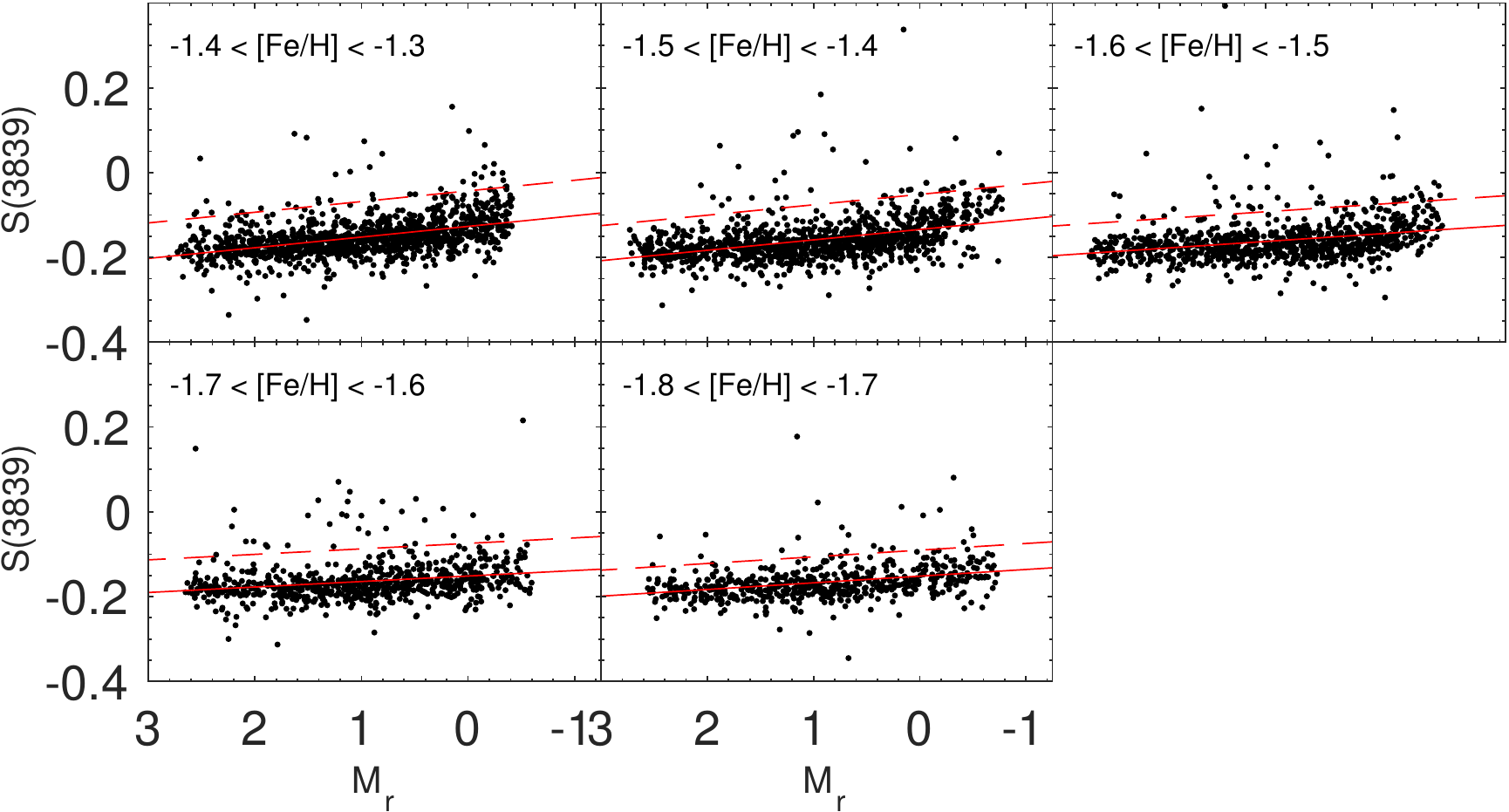}
\caption{CN-band-strength vs.\ absolute magnitude for separate
metallicity bins of 0.1 dex width. The solid line shows the best fit
to the CN-weak sequence, while the dashed red line delineates the
separation into CN-weak and CN-strong stars.}
\end{figure}

For each bin, we computed a generalized, i.e., error-weighted
histogram of $\delta$S(3839), from which we drew the distinction into
CN-weak and CN-strong stars by defining the largest separation in the
histogram, aided by a Sobel edge-detection filter. 
As an independent test, we also applied a heteroscedastic Kayes
Mixture Modeling algorithm \citep{Ashman1994} to every metallicity
subsample, which assigns probabilities to each star to belong to
either of the CN-strong or CN-weak populations. These probabilities are explicitly
assumed to be Gaussian. The resulting distribution (dashed lines in
Fig.~5) shows a vast overlap of CN-weak and CN-strong stars and leads to
a large, overall ratio of these two populations that is 
higher than previous estimates in the literature.  Considering this
overlap and the fact that our first method takes into account the more
realistic error distribution of our data, we will not pursue the KMM
separation any further.
\begin{figure}[htb]
\centering
\includegraphics[width=1\hsize]{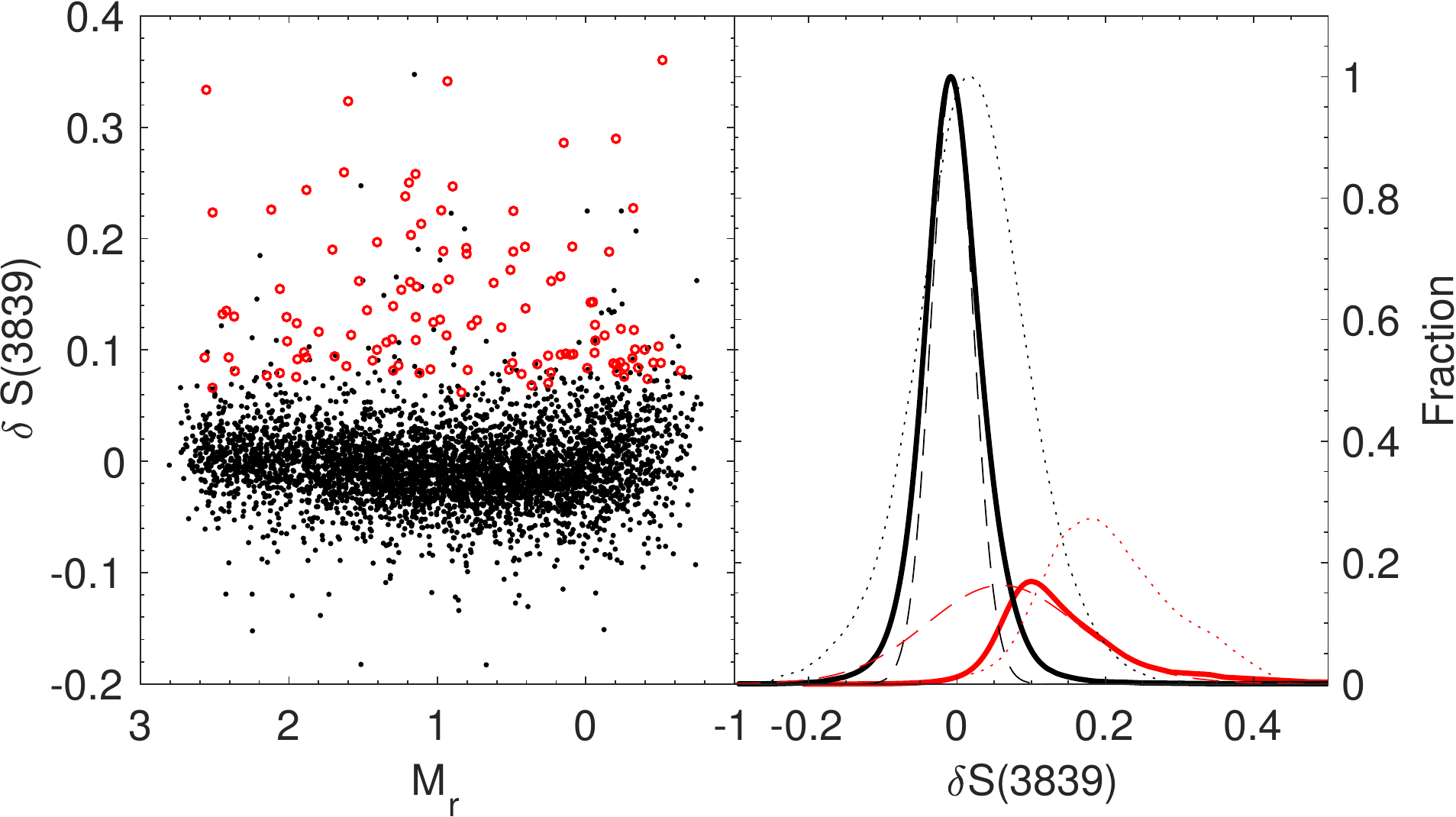}
\caption{Combined sample of CN-strong (red) and CN-normal stars. The
right panel shows the generalized histogram of both poulations, where
the curve for the CN-strong stars has been scaled up by a factor of ten.
Dotted lines show the distribution derived by \citet{Martell2010} and
the dashed curves are the results from our KMM separation.}
\end{figure}

For the construction of the final sample, we took advantage of the
fact that stellar evolution in the progenitors, whether massive or
not, produces the phenomenologically observed anticorrelations
between CN and CH band strength in GC stars by efficiently converting C 
into N.
For the present purpose this means that strong CN-bands should be
accompanied by considerably weakened CH-band strengths. 
{ Following \citet{Martell2011}, we %thus adopt all CN-strong objects more
%metal-rich than $-1.3$ dex as {\em bona-fide} GC stars and 
adopt the mean S(CH) of stars more metal-rich than $-1.3$ dex 
as a delimiter between CH-normal and CH-weak for the low-metallicity bins
that make up our present sample.} 
This selection is again carried out in metallicity bins of 0.1 dex and
indicated in Fig.~6. 
\begin{figure}[h!]
\centering
\includegraphics[width=1\hsize]{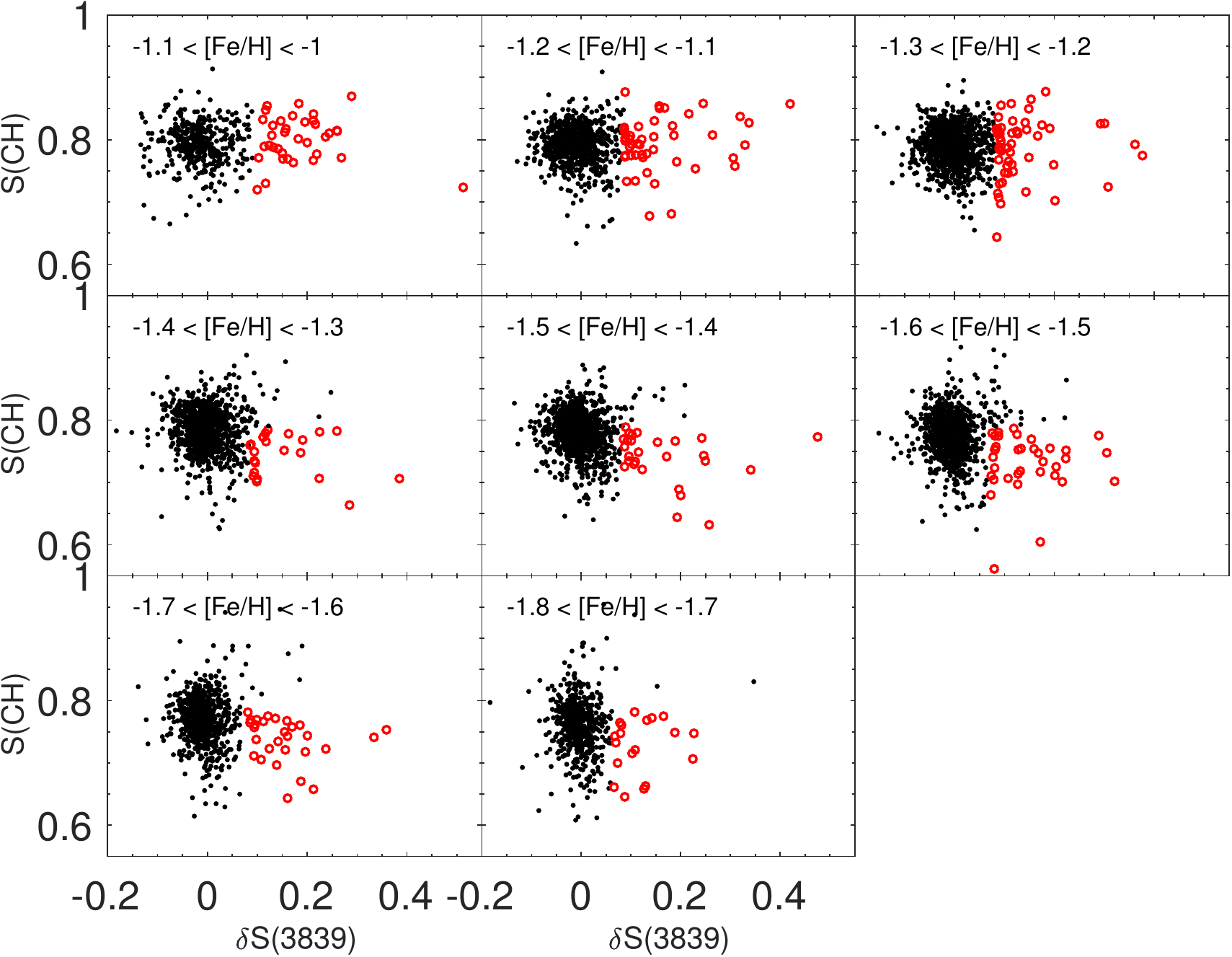}
\caption{Separation of regular halo stars that are CN-normal and
CH-strong (black dots) and the {\em bona fide} CN-strong, CH-weak
stars that we assume to have originated in now-dissolved GCs (red
circles). In concordance with Fig.~3, this was done in separate
metallicity bins. { While not used in the actual statistics, we include
in this figure stars above [Fe/H]=$-$1.3 for illustration.}}
\end{figure}

Fig.~5  displays the combined sample in the corrected
$\delta$S(3839) index, also accounting for its error distribution
(right panel).  
Owing to the additional separation criterion in terms of metallicity (Fig.~4), some of the 
CN strong stars are also CH-strong, while we do not consider them as GC-descendants. Nonetheless, 
this leads to an overlap in those Figures focusing on CN-strength alone (Figs.~4,5).

A clear bimodality of CN-strong and CN-normal stars is
seen, as is commonly found in Galactic and extragalactic GCs
\citep[e.g.,][]{Kayser2008,Larsen2018,Hollyhead2018}. 
These separate populations will be the base for all our subsequent
statistics and number counts.  Sample spectra from either group are
depicted in Fig.~7, { including those of some rejected, metal-rich examples}, 
=demonstrating the success in isolating the relevant
stellar tracers based on our spectral index analysis.
\begin{figure}[htb]
\centering
\includegraphics[width=1\hsize]{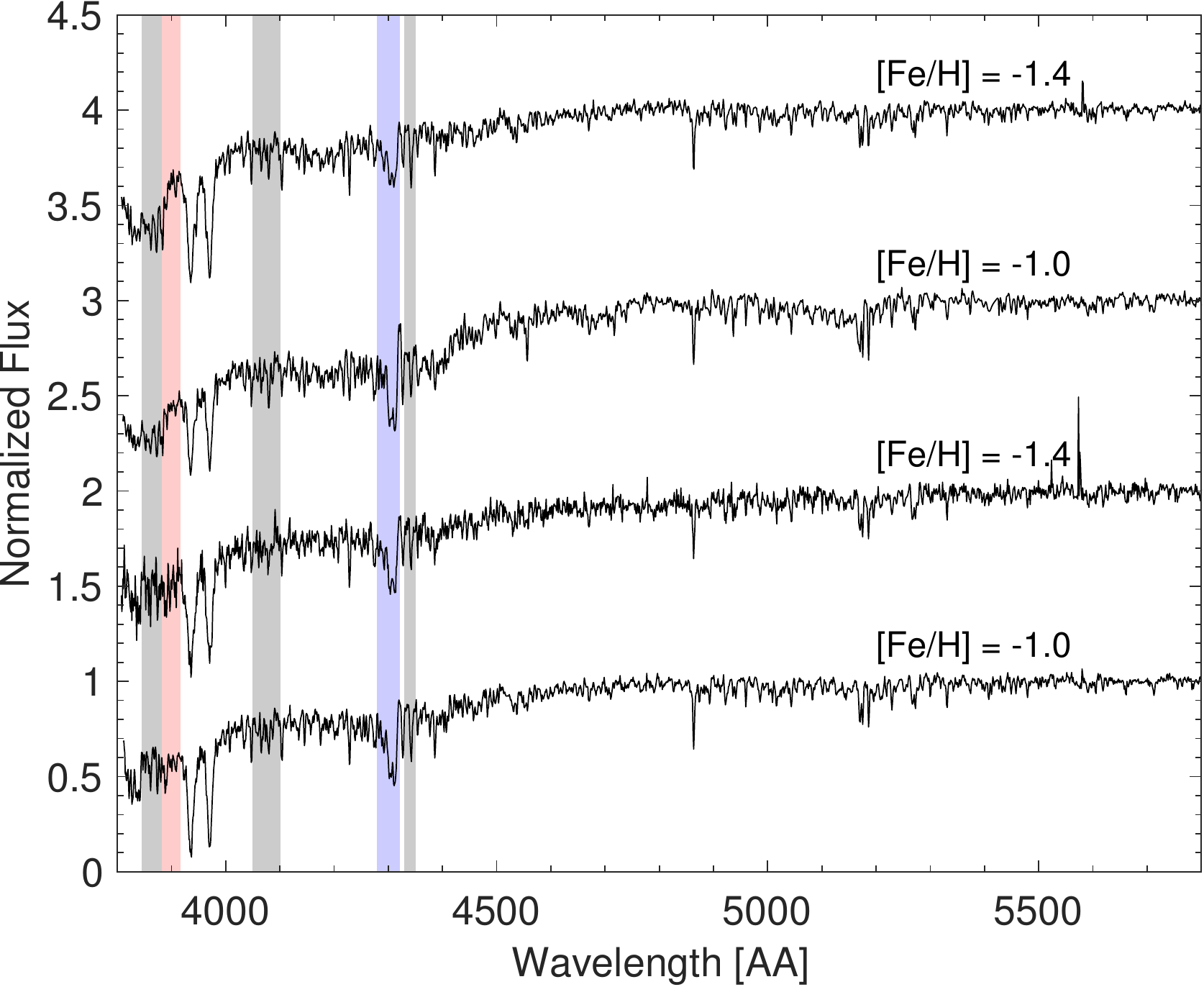}
\caption{{ Sample spectra 
for the stellar types relevant to this work. These
are (from top to bottom):  a metal-poor CN-strong star, a more metal-rich CN-strong star, 
and two CN-normal stars with stellar parameters as the former ones. Metal-rich stars above [Fe/H]=$-$1.3 were rejected
from our sample, though, given the risk of mis-classifications.} The red-
(blue-) and gray-shaded areas indicate the line and continuum
bandpasses for the S(3839) and S(CH) index after \citet{Norris1981}
and \citet{Martell2008}, see Eqs.~1,4.}
\end{figure}
\section{Halo fractions of globular cluster stars}
We find that 118  
out of our 4470  
halo giant stars are CN-strong and
CH-weak, thereby qualifying as second-generation progeny of dissolved
GCs. We note that we recover the 65 CN-strong candidates of
\citet{Martell2011} as well, which is not surprising given the same selection criteria in this work. 
This corresponds to an observed
fraction of halo stars with second-generation abundance patterns of
$f_h^{2G} = (2.6\pm0.2)$\%, where the error is solely based on Poisson
statistics. 
{ We also note that this number would increase to 3.7\% if we were to include the entire
metallicity range up to $-$1 dex, thereby yielding 252 CN-strong stars. This metallicity dependence of 
this fraction is illustrated in Fig.~8.}
\begin{figure}[htb]
\centering
\includegraphics[width=1\hsize]{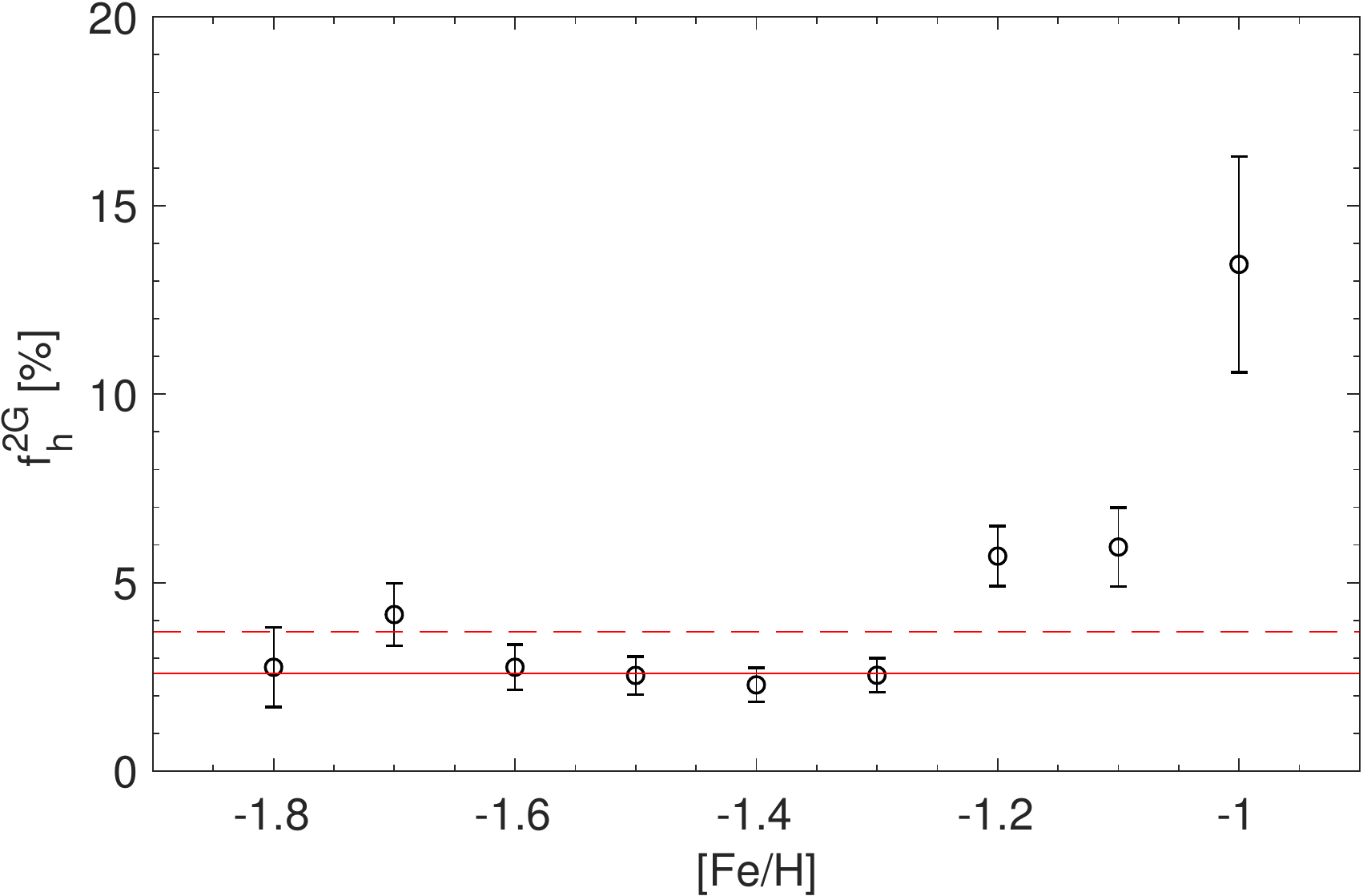}
\caption{Dependence of the observed fraction of CN-strong halo stars on metallicity. The solid line shows
our recovered, overall fraction based on the sample within $-1.8\le$[Fe/H]$\le-1.3$, whereas the dashed line
also accounts for the metal-rich tail up to $-1$ dex.}
\end{figure}
\subsection{Formalism}
Our goal is to compute the fraction of the halo that was contributed
by disrupting GCs, $f_h^{\rm GC}$. This is simply the ratio of the
total mass in former globular cluster stars in the halo
field, $M_h^{\rm GC}$, to
the total present-day stellar mass of the halo, which we adopt here as
$M_{\rm h,tot}=10^9$ M$_{\odot}$\footnote{Strictly, this number should be scaled down by a factor
of four due to the smaller footprint of the SDSS used in this work
compared to the entire halo volume.
Likewise, a  correction factor for our limited distance coverage could
be envisioned, which, however, would counteract our investigation of
the distance dependence addressed in Sect.~5.3.}.  
{We also note that the MW GC system {\em currently} accounts for about 2--3\% of the stellar halo mass \citep{Forbes2018}.}

We will use the following terminology and formalisms,
which  differ in several details from those of
\citet{Martell2010,Martell2011}. 
In our work, ``mass loss at early time'' refers to the phase of strong
loss of first-generation stars from GCs that was proposed to solve the
mass budget problem, i.e., the observational evidence that the present
stars with primordial signatures (and the respective higher mass stars
that have already perished) are insufficient to have produced the
amount of light elements required to enrich the second generation stars to the observed
levels \citep[e.g.,][]{BastianLardo2015}.  During this period, a
fraction of $f_{1G}^{\rm lost}$ of mass in the initial globular
cluster system is lost. We will discuss the implication of the exact
value of this parameter in Sect.~5.1.3 below. 
Accordingly, $M_{1G}^{\rm lost}$ is the mass lost from the globular
cluster system at these early times.
In contrast, ``late mass loss'' is taken as stars escaping from
globular clusters after the second generation has formed, whether
through internal or external processes.
\subsubsection{Ratio of first- to second generation stars}
Turning to the present-day GCs, the ratio of
first-to-second-generation stars is still debated.  Early studies,
observationally and theoretically, commonly assigned the GCs with an
equal proportion of primordial and enriched populations
\citep[e.g.,][]{DErcole2008,Carretta2009NaO}.  Based on their study of
33 MW CGs \citet{BastianLardo2015} concluded that the fraction of
second-generation stars is 68\% and independent of cluster parameters
such as mass, metallicity, or location within the Galaxy. 
Conversely, \citet{Milone2017} found a clear trend of the second
population with present-day GC mass in the form of a decreasing
occurrence of the enriched stars with decreasing mass \citep[see
also][]{Gerber2018}. Moreover, the population fractions are expected
to vary with distance from each GC's center
\citep[e.g.,][]{Milone2009,Vanderbeke2015,Nardiello2018} and change
during the clusters' dynamical evolution \citep{Vesperini2013}.  The
influence of varying this ratio on the final derived fraction of the
halo stemming from dissolved GCs is illustrated in Fig.~9.  
{Finally, \citet{Baumgardt2017} do not find a correlation between
the first-generation fraction and the global mass function of GCs, but
instead a trend of the escape velocity with the first-generation
fraction, which reflects the finding of \citet{Milone2017} of a mass-dependence.}
\begin{figure}[t!]
\centering
\includegraphics[width=.97\hsize]{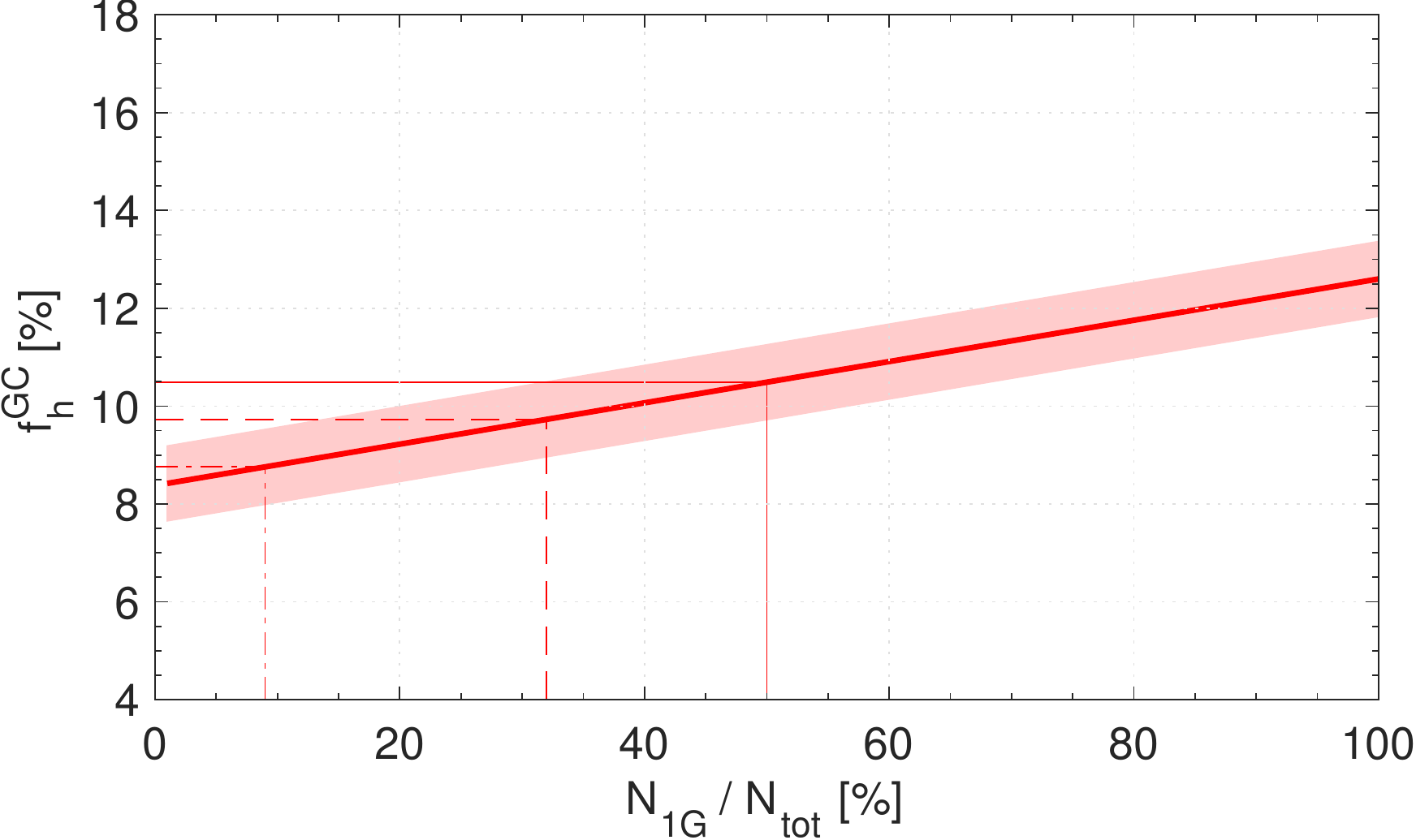}
\includegraphics[width=.97\hsize]{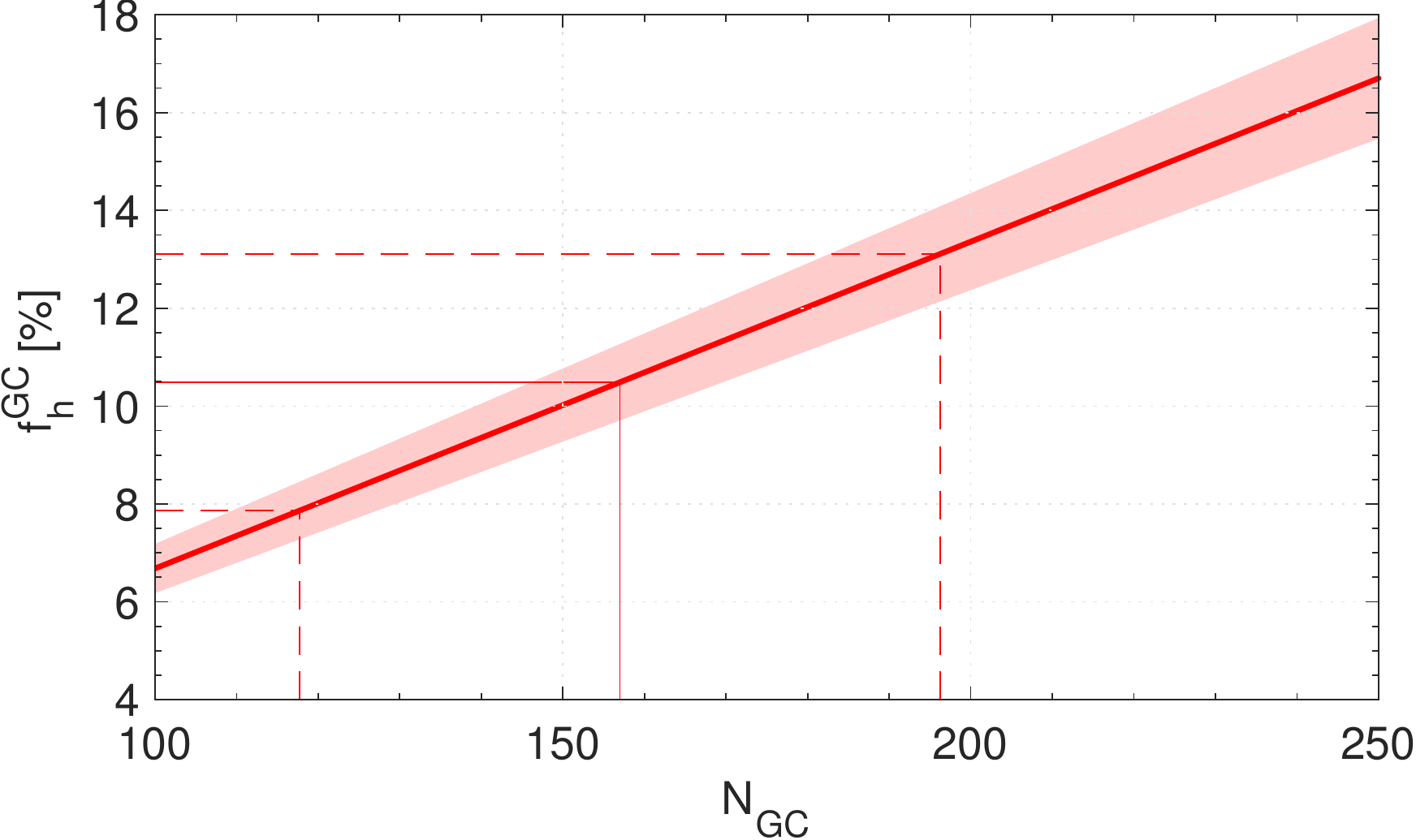}
\includegraphics[width=0.94\hsize]{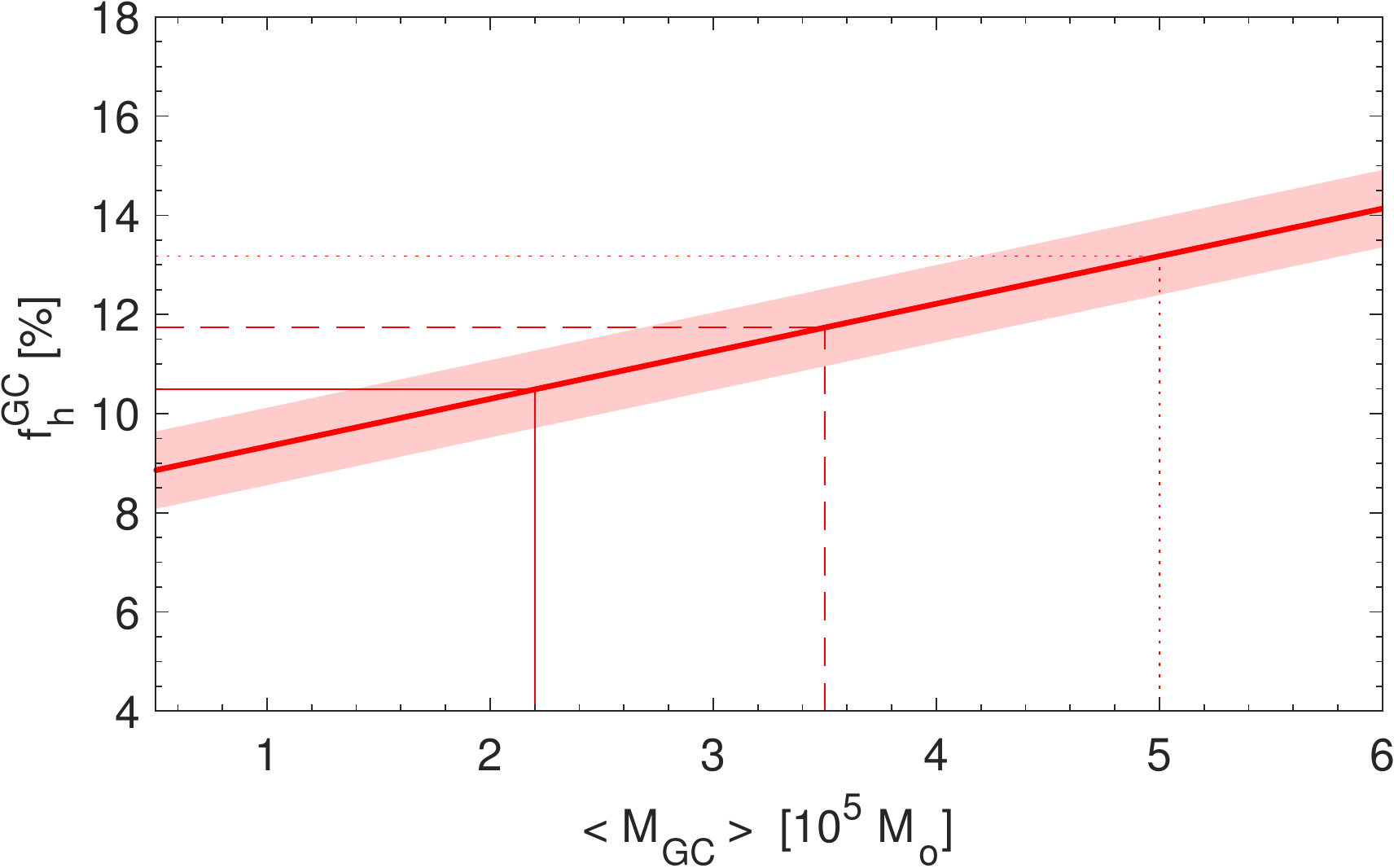}
\caption{{\em Top panel}: Dependence of the final fraction of donated
halo stars on the adopted fraction of first-generation stars.  The
solid, dashed, and dashed-dotted lines are for ratios of 0.5
\citep{Carretta2009NaO}, 0.32 \citep{BastianLardo2015}, and 0.09, the
minimum value in the catalog of \citet{Milone2017}. This calculation
adopted a ratio of early mass loss fraction (Sect. 5.1.3) of 
{56\%}
 and an average, present-day GC mass of $2.2\times10^5$ M$_{\odot}$.
{\em Middle panel:} Influence of the number of present-day GCs in our
analysis. Here, the solid line shows the 157 objects in the
\citet{Harris1996} catalog, while the dashed lines correspond to a
variation by $\pm$25\%, e.g., due to selecting only halo GCs.
{\em Bottom panel:} Dependence on the adopted, mean present-day GC
mass.  Red lines show the masses based on the \citet{Harris1996}
catalog adopted in this work (solid), the catalog of GC masses from
\citet[][dashed]{McLaughlin2005}, and a mass of $5\times10^5$
M$_{\odot}$ (dotted) as adopted by \citet{Martell2011}. Here, the calculations
adopted a ratio of first-to-second-generation stars of 50:50 and an
early mass loss rate of 
{56\%.}}
\end{figure}

In principle, one can also assume a realistic distribution of GC
masses (see Sect.~5.1.2 below) and attach a CN-strong fraction to each
mass value via the functional trend implied by \citet{Milone2017}.
However, as the range of resulting fractions shown in Fig.~9 is fairly
small, we adopt for simplicity that the GCs consist of 50\%
first-generation stars and 50\% second-generation stars, which is at
the higher end of the values suggested in the literature, noting that
this will provide an upper limit in our final results. 
\subsubsection{Cluster masses and number of present-day GCs}
To date, $N_{\rm GC}$ old GCs have been identified in the MW -- this
is the number of surviving clusters in the Galaxy, for which we adopt
a value of 157,  corresponding to the number of known GCs in the
\citet[][2010 edition]{Harris1996} catalog.  As we are dealing here
with the build-up of the stellar {\em halo}, by rights only the halo
clusters should be considered, which make up $\sim$75\% of this
database.  This would {only mildly decrease} the resulting halo-fraction obtained in
Sect.~5.3. from {11 to 8\%.}  The influence of this parameter on our
final result is shown in the middle panel of Fig.~9. 

In order to convert this number into an overall mass of the Galactic
GC system, we consulted the catalog of known MW GCs by
\citet{Harris1996} and computed cluster luminosities from their
absolute magnitudes. For their stellar populations, we assume a
stellar mass-to-light ratio of $M/L\,=\,2\,(M/L)_{\odot}$, noting that this
may vary to within a factor of about two
\citep{Illingworth1976,Pryor1993,Strader2009}.  Summing up these
values results in a total mass of the GC system of $M_{\rm tot} =
3.5\times10^7$ M$_{\odot}$.  In turn, this yields a simplistic
estimate of the mean mass of a typical, present-day MW GC of
$<$$M_{\rm GC}$$>$ = $M_{\rm tot} / N_{\rm GC} = 2.2\times10^5$
M$_{\odot}$.

Alternatively, \citet{McLaughlin2005} determined dynamical masses and
mass-to-light ratios of approximately half of the MW's GCs and found
$M/L$ ratios that are consistent with the value of 2 adopted in our
work. The average GC mass in their catalog is higher by $\sim$60\%. 
However, this has only a minor influence during the propagation
through our formalism below and the final, accreted halo fraction does
not change by more than two percentage points.  The influence of varying the
present-day, mean GC mass on the final derived fraction of the halo
stemming from dissolved GCs is illustrated in the bottom panel of
Fig.~9.

Now we designate $N_{\rm Diss}$ and $M_{\rm Diss}$ as the number of
completely dissolved GCs and the stellar mass lost through complete
cluster dissolution, respectively, which are needed to explain the
observed fraction of second-generation halo stars ($f_h^{2G}$).
Thus we can, firstly, determine the mass that has been lost at early
times as
\begin{equation}
M_{1G}^{\rm \, lost}\,=\,\frac{M_{\rm tot}}{2}\left( \frac{1}{1-f_{1G}^{\rm \, lost}} - 1\right)\,=\,\frac{M_{\rm tot} \, f_{1G}^{\rm \, lost}}{2\,\left(1-f_{1G}^{\rm \, lost}\right)}.
\end{equation}

Clusters that dissolve completely contribute, in turn, via their early
mass loss, as well as through their final masses, thus:
\begin{eqnarray}
M_{Diss}\,&=&\,\frac{N_{\rm Diss}}{N_{\rm GC}}\,\left( \, M_{\rm tot} \,+\,  \frac{M_{\rm tot} \, f_{1G}^{\rm \, lost}}{2\,\left(1-f_{1G}^{\rm \, lost}\right)} \, \right) \\
\,&=&\,\frac{N_{\rm Diss}\,M_{\rm tot}}{2\,N_{\rm GC}}\,\frac{2-f_{1G}^{\rm \, lost}}{1-f_{1G}^{\rm \, lost}}.
\end{eqnarray}

Thus the total mass of GC stars in the halo field can be expressed by
summing the above as
\begin{eqnarray}
M_h^{\rm GC}\,&=&\,M_{1G}^{\rm \, lost}\,+\,M_{\rm Diss} \\
\,&=&\,\frac{M_{\rm tot}}{2\,\left(\,1-f_{1G}^{\rm \, lost}\,\right)}\,\left(\,f_{1G}^{\rm \, lost}\,+\,\frac{N_{\rm Diss}}{N_{\rm GC}} \, \left(2-f_{1G}^{\rm \, lost}\right) \, \right). 
\end{eqnarray}
\subsubsection{Mass loss from the first generation}
Previous studies have adopted a large fraction for $f_{1G}^{\rm \,
lost}$ of up to 90\% early cluster mass loss, preferentially if the
clusters were massive.  This is based, e.g., on the AGB scenario
\citep{DErcole2008,Conroy2012}. 
This accounts for cluster evolutionary effects such as tidal
disruption, two-body relaxation, energy input from supernovae, or
residual gas loss \citep[e.g.,][]{Gnedin1997}, but not effects like
``infant mortality'', i.e., the disruption of young clusters already
during their formation stages \citep[cf.][]{Bastian2006}.
Likewise, in a scenario where fast rotating, massive stars were the
main polluters of the second generation, models predict that the
clusters were once more massive by up to factor of 25 ($f_{1G}^{\rm \,
lost}=0.96$).  However, recent models and observations do not further
support such a strong mass loss.  For instance, observations of GC
populations in dwarf galaxies \citep{Larsen2012,Larsen2014} and young
star clusters in merging galaxies \citep{CabreraZiri2015} ascertain
lower mass-loss fractions of $f_{1G}^{\rm lost}\le$80\%. 

Likewise,  \citet{BaumgardtSollima2017} argued for a loss fraction for
a ``typical GC'' of 75\%, which could, however, reach as much as 90\%
for the most massive objects, based on $N$-body simulations in
comparison with the observed mass functions of 35 Galactic GCs.  Also
the simulations of \citet{Webb2015}, accounting for orbital evolution,
indicate that an average cluster was initially 4.5 times more massive
($f_{1G}^{\rm \, lost}=0.78$). 
We also note that 
\citet{Kruijssen2015} 
considered
a  two-phase model
for GC evolution, according to which only the least massive and/or
metal-rich clusters suffered from high mass loss rates, whereas an
average system has experienced typically $f_{1G}^{\rm lost}\sim 2/3$.

It stands to further reason that in the mass loss considered above, only first generation stars are lost, not those from the second.  
While originally hypothesised to solve the mass budget problem,  a number of works 
have  since shown that this heavy mass loss of only first generation stars encounters a number of problems
 (see the discussion in \citealt{Bastian2018}).  A strong argument for this  is the result from 
\citet{Milone2017} that the fraction of enriched stars increases with cluster mass, 
whereas in the heavy mass-loss scenario the opposite trend is expected.
 This suggests that 
clusters did not undergo very strong mass loss 
 and that the current fraction is close to the initial fraction -- GCs  
did not preferentially lose large fractions of first generation stars. \citet{ReinaCampos2018}
 found that clusters with 
present-day masses in excess of 10$^5$ M$_{\odot}$ once were only more
massive at birth (after accounting for stellar evolution) by factors of 2--4.  
For our purpose this would imply that, if second generation stars are lost at a similar rate to those from the first population, then 
one would only need to correct for the relative population fractions (i.e., 50/50), and not any 
additional factors. For completeness, we chose to follow our previous assumptions and to retain the full correction factors, referring again 
to Fig.~9 for a quantitative estimate of their variations.
{Further discussions of the Initial Mass Function and its relevance for GC evolution can be found
in \citet{Forbes2018}.}

\subsubsection{Mass loss by stellar evolution}
{Many of the GC models discussed in Sect.~5.1.3 account for stellar
evolution, where about a factor of two comes from stellar evolution
and the other half stems from the aforementioned dynamical mass loss,
viz., either two-body relaxation or tidal shocking. 
This equally affects stars that are part of a cluster and those that are lost from the outer regions. 

Most importantly, our empirical analysis hinges on the comparison of 
halo stars with GCs of similar age and metallicity. 
If we assert that the clusters were two times more massive at birth than
now, due to stellar evolution, we must also recognize that  also the halo was twice as 
massive due to the same channels of stellar evolution.
Consequentially, the mass loss factors introduced in Sect.~5.1.3 need  
be lowered by a factor of two. In the following, we will start with the assumption that the initial GCs were
4.5 times as massive as they are today, as argued by \citet{Webb2015}, who did include
stellar evolution. Therefore, we halve this value to 2.25 times 
larger initial masses, neglecting stellar evolution, which corresponds to our final, adopted 
fraction of 56\% for the mass loss from the first generation.

In Fig.~10 we show the resulting mass fraction of the
halo from GC disruption ($f_h^{\rm GC}$), as per the calculations
concluded in Sect.~5.2, as a function of the adopted mass-loss rate
$f_{1G}^{\rm lost}$. 
Whether the cause lies in the details of the model or 
in factoring in the effect of stellar evolution is redundant in this figure, 
as it merely shows the dependence of our results with respect to the 
quantitative assumption. To this end, we rather highlight three sets of adopted model fractions, each shown for 
the cases of inclusion and exclusion of stellar evolution.}
\begin{figure}[tb]
\centering
\includegraphics[width=1\hsize]{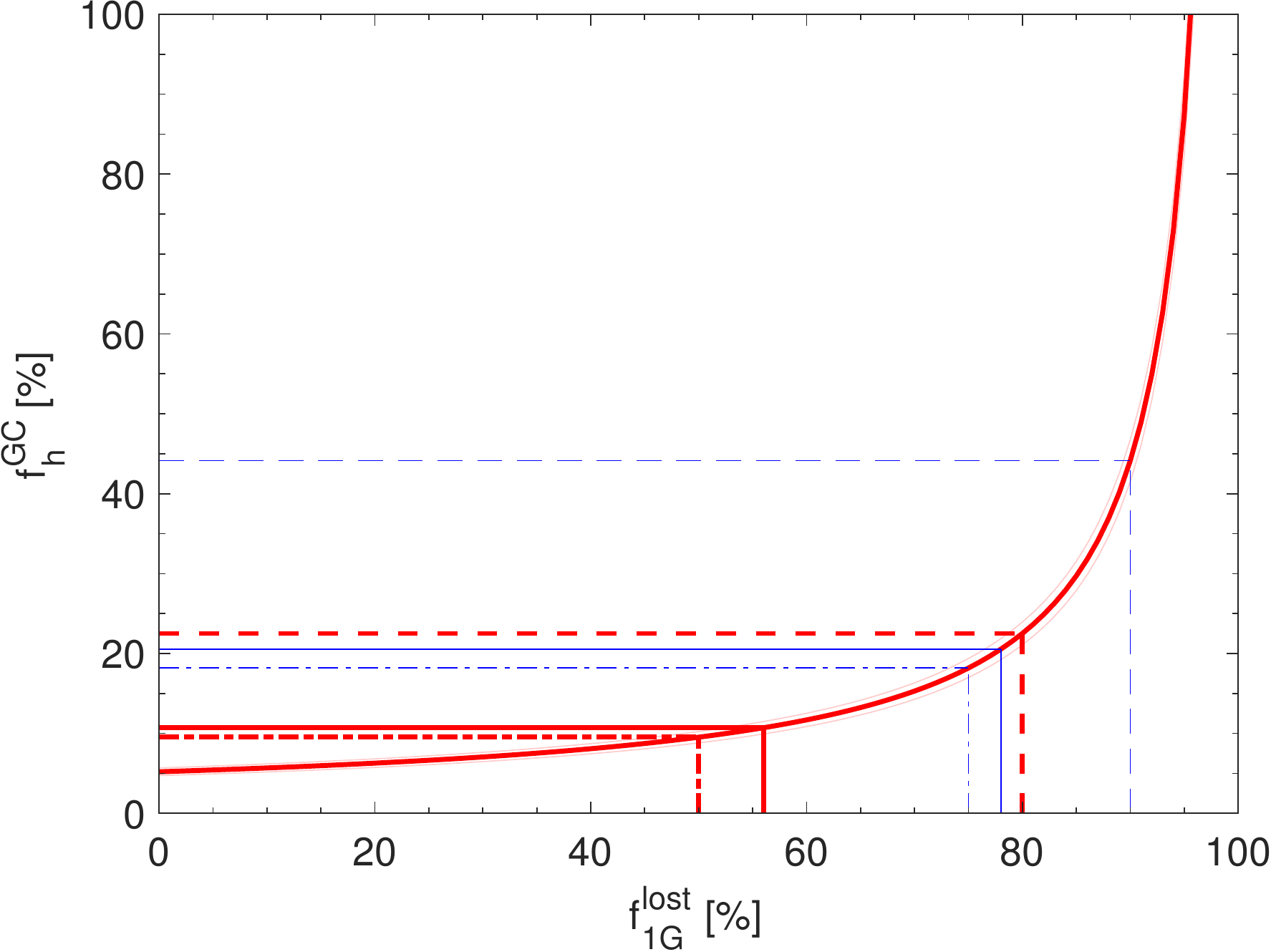}
\caption{Dependence of the final fraction of accreted halo stars on
the adopted mass-loss rate of first-generation stars.  The lines indicate
the fractions adopted by various works:
\citet[][dash-dotted]{BaumgardtSollima2017},
\citet[][dashed]{Martell2011}, and our present study (solid). 
{Here, red lines show the cases that neglect stellar evolution, while blue lines show values where this is accounted for.}
This calculation adopted a ratio of first-to-second-generation stars of
50:50 and an average, present-day GC mass of $2.2\times10^5$
M$_{\odot}$.}
\end{figure}

Even in the hypothetical absence of any such
early losses (i.e., Eqs.~9,10 with $f_{1G}^{\rm lost}$=0), one would
still end up with a theoretical lower limit of $2\,f_h^{2G}$={ 5\%} GC
contributions to the present-day stellar halo. This is due to those
clusters that have by now fully dissolved and that also contributed to
the halo build-up via later mass loss throughout their lives. 
{ In fact, in the MW halo, a mass loss fraction of $f_{1G}^{\rm lost}$=0 
is certainly conceivable; we note that this does not imply that there was no mass loss from any GC, but 
rather means that no stars of the first generation were lost, or that 
the fraction of first and second generation stars lost was roughly equal.}
\subsection{Resulting halo fraction}
At this point, we will consider the actual number of dissolved GCs,
$N_{\rm Diss}$, that are required to provide the fraction of
second-generation stars, $f_h^{2G}$, that  we found in the Galactic
halo in the previous sections.  As elaborated in \citet{Martell2011},
this value is given as 
\begin{equation}
N_{\rm Diss}\,=\,\frac{2\,f_h^{2G}\,M_{\rm h,tot}}{<M_{\rm GC}>}
\end{equation}
with the factor two stemming from the adopted similar fractions of
first- and second-generation stars in the early clusters. 
{ It is this factor, which is not strongly constrained by observations, 
that dominates the final inferred fraction. Therefore 
we highlighted its influence in Fig.~9 (top); see Sect.~5.1.1.}
Based on our 
{ observed fraction of
$f_h^{2G}=0.026\pm0.002$} and the above assumptions, we obtain a large
{ number of 240$\pm$22 clusters} to be dissolved.  This is more than
{ twice } the value found by \citet{Martell2011}, but 
 their working hypothesis had been a higher average
cluster mass of 5$\times10^5$ M$_{\odot}$.

As a result, we find that the fraction of the stellar halo that
originates from disrupted GCs, $f_h^{\rm GC}=M_h^{\rm GC} / M_{\rm
h,tot} = $ {0.11$\pm$0.01}, under the assumption of a mass loss fraction
(Sect.~ 5.2) of {56\%}. 
The influence of different mass loss prescriptions is shown in Fig.~10 and has been discussed before in Sects.~5.1.3 and 5.1.4.
\citet{Martell2010} stated that a high fraction of up to 50\% of the
halo field must have formed in massive star clusters, constituting the
CN-strong population observed nowadays, with a further, unknown
contribution of CN-weak stars from now fully dissolved systems.
Expanding their analysis using larger data sets, and using their
finding of a lower $f_h^{2G}$ of 2.5\%, \citet{Martell2011} estimated
a lower limit of 17\% of halo stars with both first- and
second-generation element patterns, under the omission of low-mass
clusters that are long gone. 
The observed fraction of GC-like halo stars from the latter work 
is consistent with the value found in our present work, {while our inferred 
value for donated halo stars is slightly smaller}. This is, however, 
chiefly due to the updated statistical treatment and
different model assumptions, which we amply addressed through our Figs.~9 and 10.

 As a further test, as mentioned in Sect.~5.1.1, we 
 assigned to each present-day GC mass in the compilation of \citet{McLaughlin2005}
  a corresponding first-to-second generation fraction using the empirical trend from 
 \citet{Milone2017} rather than adopting the generic 50/50 ratio. 
 As a result, we would find $N_{\rm Diss}$=100$\pm$8 dissolved clusters
 and a donated-halo fraction of {6\%}. This majorly hinges on the assumption of the 
 present-day mass distribution. Any sophisticated treatment would need to adopt a lower mass limit above which to define a ``GC'', 
 initial population fractions \'a la \citet{ReinaCampos2018}, and/or explicit initial mass functions,  the latter being highly conjectural 
 at this point, so that we continue our discussion with our above finding of $f_h^{\rm GC}$ = {0.11.}
 
We note that all our results are merely upper limits, since we
assumed here that {\em all} halo stars with second-generation
chemical imprints are the result of cluster dissolution. Alternative
sources for these chemical patterns could be the effects of
post-mass-transfer AGB binaries \citep[e.g.,][]{Pols2012} or internal
mixing processes during stellar evolution within the stars themselves
\citep{Spite2005,Stancliffe2009,CJHansen2016}.
\subsection{Dependence on distance}
Fig.~11  shows the trend of the CN-strong stars in our sample as a
function of Galactocentric distance; for the latter we adopted a
distance to the Galactic center of 8.34 kpc \citep{Reid2014}.  
Both the observed fraction of stars with second-generation chemistry
(middle panel) and the derived fraction of that portion of the halo that
would stem from GCs (bottom panel) show a 
{ predominantly flat 
trend within
$\sim$30 kpc}, while there is an apparent rise beyond this radius,
albeit hampered by the overall small-number statistics and accordingly large error-bars 
at these large distances.   

We further caution that the conversion between these fractions,
following Eqs. 5--10, assumes that the distribution of the formerly
disrupted GCs (entering via $N_{\rm Diss}$) with Galactocentric
radius was isotropic.  This also presumes that the orbital history of
the entire MW GC system is considered irrelevant in that we reference
our formalism to the present-day, total number of GCs.  In fact, the
MW GCs are rather distributed anisotropically and separated into the
underlying Galactic components such as disks, bulge, or halo GCs.  
This is, however, not crucial to the present analysis as we are mainly concerned with halo GCs and halo stars as per our
selection criteria.
\begin{figure}[tb]
\centering
\includegraphics[width=1\hsize]{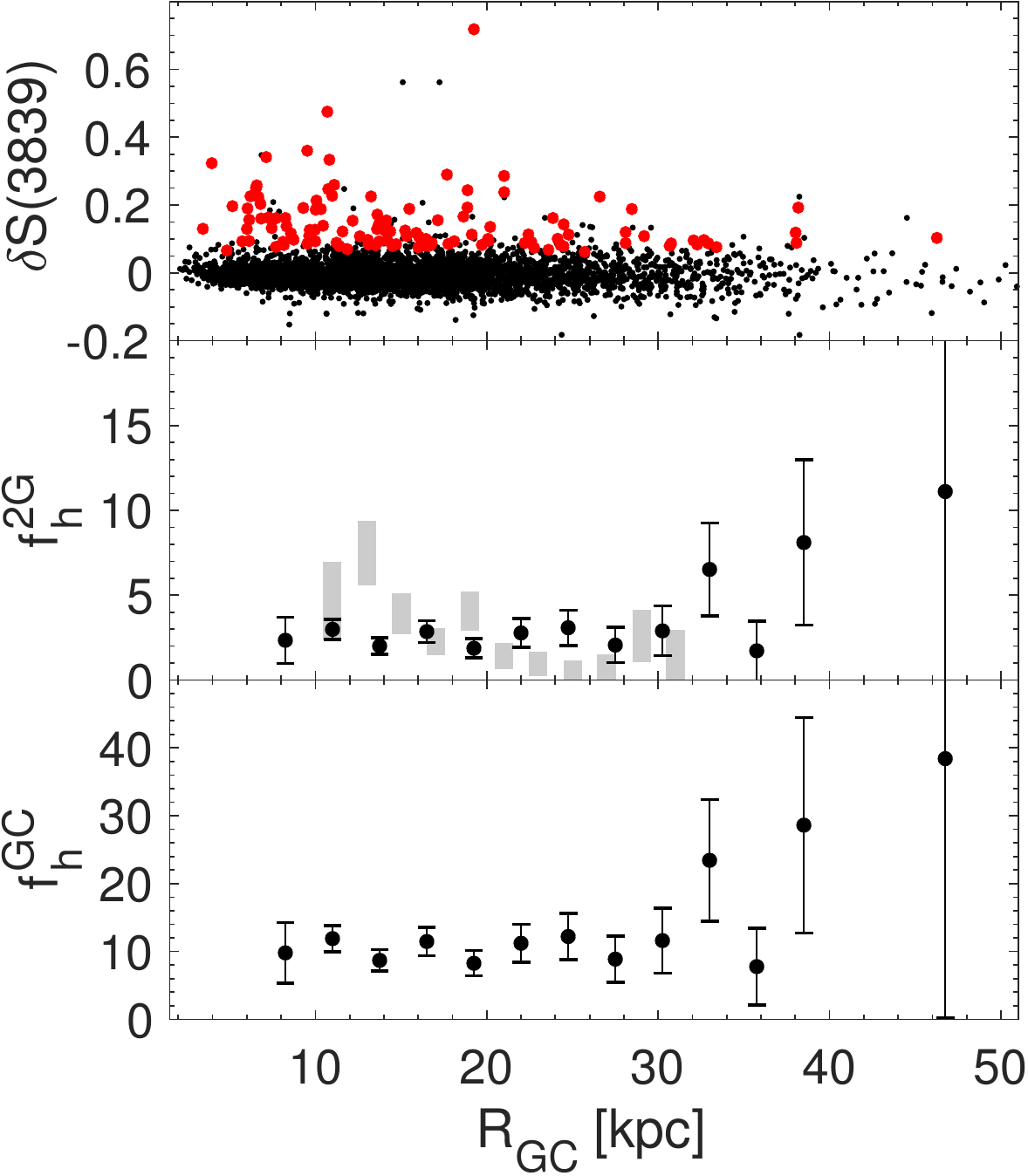}
\caption{Top panel: distribution of CN-weak (black) and CN-strong (red)
stars with Galactocentric distance.  Middle and bottom panels:
resulting fraction of CN-strong to CN-weak stars as function of
Galactocentric distance. The errorbars are based on Poisson
statistics.  Gray-shaded areas show the radial trend of the 65
CN-strong stars found by \citet{Martell2011}.}
\end{figure}

The radial trend of Fig.~11 is, {  qualitatively, reminiscent of those seen in other studies 
\citep{Carollo2013} and other halo tracers} such as carbon stars
\citep{Carollo2012}, RR Lyrae \citep{Medina2018}, and similarly for
the radial (metallicity) profiles in the neighbouring Andromeda galaxy
M31 \citep{Ibata2005,Koch2008M31} and other external galaxies
\citep[e.g.,][]{Forbes2011}, { albeit the latter examples 
show generally stronger declines with radius within their inner halos}. 

\citet{Martell2011} found one CN-strong star beyond 30 kpc, giving
rise to an apparent bump in their distance distribution of the
CN-strong- to -normal fraction.  Such a feature, though not
significant is also seen in our profile at 30--35 kpc, which could
indicate the presence of a real substructure in the halo at these far
distances.  Overall, we find { 13} second-generation GC stars at
distances beyond 30 kpc, with the farthest CN-strong star in our
sample lying at { 47$\pm$6 kpc}.
\section{Discussion}
Our exploitation of the large-scale, low-resolution spectroscopic
surveys of the SDSS has resulted in our detection of more than a hundred
stars in the halo field that show strong CN- and weak CH-bands,
which are the key signatures of second-generation GC stars.  We thus
estimate that about one {tenth} of the present-day MW halo originates from
disrupted GC stars; this value lies {below} previous estimates that
were based on SDSS data from earlier releases with  smaller numbers of selected stars {and that 
used different formalisms}. 

Our estimate hinges on several assumptions, some of which pose significant limitations to the extractability of a halo fraction from 
observations of stars with GC-like chemistry. Amongst these are the:

\vspace{1ex}
\noindent
{\em Ratio of first-to-second generations (Sect. 5.1.1)}

\vspace{1ex}
\noindent
The dependence of this parameter on global GC parameters such as mass
(be it initial or present-day) or metallicity is still not fully
settled.  We have adopted here a value of 50 per cent of
first-generation stars, which is an upper limit to the values
suggested by theories and observations. 
{ Our observational finding that second generation stars make up 2.5\% of the stellar halo could, effectively, be doubled 
when correcting for the population ratio. 
If some mechanism were only to remove the first generation, then our derived fraction would only be a lower limit, with a final number 
between 5\% and 100\% depending on the assumed mass loss rate. However, as per our  formalism, the 
change in this value when changing the respective population ratio by a few tens of per cent is marginal.}

\vspace{1ex}
\noindent
{\em Mass loss at early times (Sects. 5.1.3, 5.1.4.)}

\vspace{1ex}
\noindent
{Here, we have adopted a mass loss fraction of 56\%.
 While substantially stronger} mass loss is suggested by several theories
of GC enrichment and evolution, this would, as per our formalism,
imply  that, unrealistically, the entire MW halo {\em or more}
consists of stars formerly hosted by now-disrupted GCs.  This problem
of ``overpopulation'' has also been noted for the MW bulge
\citep{Schiavon2017}, in the context of mismatches in the metallicity
distribution function of the Fornax dwarf spheroidal galaxy's GC
population when scaled to that of its field stars \citep{Larsen2012},
and theoretically addressed by \citet{Vesperini2010}.
{We restate that our adopted value assumes that stellar evolution 
is not accounted for as it affects stars irrespective of their environment (read: halo and 
GCs), which prompted the need to halve the literature value from the model adopted in our work.}  

\vspace{1ex}
\noindent
{\em Mass loss from the second generation}

\vspace{1ex}
\noindent
Canonical models of GC evolution and disruption mainly engage mass
loss at early times, as also inherent in our formalism.  However, it
is natural to assume that the second generation stars escape from the
GCs at an early stage, as suggested by \citet{Schaerer2011}. The
implication of their model (with fast rotating massive stars as the
main enrichers) is that the GCs were initially more massive by factors
of $\sim$10 times (only first generation lost) or 25 times (also
second-generation loss). In the scenario proposed by
\citet{Schaerer2011} this also affects the conclusions drawn on the
fraction that present-day GCs have contributed to the low-mass stars
in the Galactic halo: while the former mass-loss fraction results in
5--8\%, also accounting for the second generation increases this value
to the 20\% level.
Similarly, \citet{BaumgardtSollima2017} expect that a typical 
MW GC (permitting for various mass functions) would have lost about 75\% of its mass since
formation, while more evolved clusters have already lost more than
90 \%  and should dissolve over the next
1--2 Gyr, which, however, does not give us a handle on distinguishing stars of the first and second generation in the field. 
{In accordance with our discussion in Sect.~5.1.4, we note that these models did include the effects of stellar evolution}.

\vspace{1ex}
\noindent
{\em Stellar halo mass}

\vspace{1ex}
\noindent
 In our statistics we have explicitly adopted one single number
(10$^9$ M$_{\odot}$) for the mass of the stellar halo, while more
sophisticated correction factors for the footprint of the employed
surveys (here, SDSS) could be envisioned -- which could be argued for
both our target selection and the number of present-day GCs to which
we scaled our formalism.  Moreover, our approach ignored the orbital
histories of the observed MW GCs and assumed an isotropic
distribution. 

Another potentially aggravating factor is the exact value for the
stellar halo mass. On the other hand, many studies have attempted
weighing the entire halo (i.e., dark plus stellar) from a variety of
tracers
\citep[e.g.,][]{Wilkinson1999,Sakamoto2003,Xue2008,Kafle2014,Wang2015,Elias2018},
and the resulting range is less critical in the present analysis. As
the halo mass enters inversely in the fraction of the GC-built halo,
corrections can be easily implemented. 

\vspace{1ex}
\noindent
{\em Present-day GCs (Sect.~5.1.2)}

\vspace{1ex}
\noindent
In the comparison of the number of dissolved GCs to the present-day GC
population in the MW we have adopted the census of the
\citet{Harris1996} catalog. Ever since its latest revision in 2010,
more GCs are being discovered towards all major Galactic components
including the Galactic halo
\citep[e.g.,][]{Balbinot2013,Laevens2014,Kim2016,Koposov2017,Ryu2018}.
In particular, an estimated dozen objects towards low Galactic
latitudes (while not necessarily relevant to studies of the halo) are
still awaiting discovery \citep{Ivanov2005}. Since a generous decrease
in the number of present-day GCs (excluding disk and bulge objects) or
its increase (extrapolating future discoveries) by 25\% only amounts
to a change {of a few percentage points} in our results, the exact choice of this parameter is
uncritical.  

\vspace{1ex}
\noindent
{\em Low-mass clusters} 

\vspace{1ex}
\noindent
There is now obervational support for a minimum mass on the order of a
few $10^4$ M$_{\odot}$ for GCs to exhibit multiple populations and
light element variations
\citep{Simpson2017ESO,Bragaglia2017,Bastian2018}. Furthermore, some
systems appear to be first-generation only systems
\citep{Villanova2013} without any photometric or chemical evidence for
the presence of a second generation, as we are looking for in this
present work.
However, also the lowest-mass systems and those with peculiar
population mixes are likely to have experienced destruction and
contributed to the halo build-up. As a consequence, these objects
would not spawn any measurable quantity of the enriched
second-generation stars, while they could well have contributed
first-generation (CN-weak) stars which we generally discarded from
consideration.  This factor thus leads to an underestimate of our
resulting fraction of the halo stars with GC origin by an unknown
amount. 
Indeed, if one assumes that {\em all} star formation happens in a
clustered manner, we will also miss the contributions of the early
counterparts of open clusters and associations, which would not have
included second-generation stars.  Our current work refers to
contributions by GCs with masses of at least a few times $10^4$
M$_{\odot}$, which may reasonably be assumed to have formed
second-generation stars.

\vspace{1ex}

The final {\em a posteriori} value of the donated halo fraction
$f_{\rm h}^{\rm GC}$ thus relies on these, mainly model-dependent,
assumptions, which hampers a clear-cut case distinction. Conversely,
as per our formalism above, in particular through Figs.\ 9 and 10, the
parameter space of the early mass-loss fraction, GC mass, and
generation ratio could be majorly constrained if models were to
independently predict $f_{\rm h}^{\rm GC}$.  
\citet{Kruijssen2015} used his ``end-to-end (yet simple)'' model to
conclude that the minimum mass for the formation of multiple
populations was 10$^5$ M$_{\odot}$ at $z\ge2$ under the observational
constraint that about 2\% of the halo consists of the CN-strong (and
enriched in other nucleosynthetic products of the relevant burning
processes) stars targeted in this work. 
The conclusion of
\citet{Kruijssen2015} is now bolstered by our present, empirical finding of this fraction lying 
at the 2\% level. 

By excluding stellar evolutionary mass loss (occurring both in the halo and the GCs) we lowered the estimated halo fraction 
from 23\% to the 10\%-level, which would represent the case that the stars lost from GCs were 100\% from the first generation, so our result 
is clearly {a mere upper limit}. 
This also constitutes a significant difference in the interpretation with previous works, but our result it is now based on 
empirical relations instead of the ``arbitrary'', theoretical 
correction factor required by the scenarios invoking pollution by AGB  or fast rotating massive 
stars.
{We also reiterate that the early loss of first generation stars 
 is observationally unconstrained and with a larger early mass loss rate, 
 the fraction could still be higher than 11\%.}

Our findings may provide qualitative evidence for the inner/outer halo
dichotomy, which is a consequence of hierarchical galaxy formation
where an outer halo is dominated by an ex-situ, accreted component,
whereas the inner halo results from both in-situ formation and the
accretion of a small number of more massive satellites
\citep{Cooper2013,Pillepich2015}, but we no longer see the pronounced
decline in the fraction of stars with a GC origin towards the outer
reaches of the inner halo found in \citet{Martell2011}. 

Cosmological simulations are concerned with the accretion of
dark-matter dominated sub-galactic fragments related to progenitors of
present-day dwarf spheroidal and dwarf irregular galaxies (which, in
turn, can bring in their own GC systems; e.g.,
\citealt{Larsen2001,Law2010,Hendricks2016,Kruijssen2018,Myeong2018,Helmi2018GaiaEnceladus}).
Detecting the traces of disrupted GCs (which may include both objects
formed in-situ as well as accreted GCs) as a main purveyor of the
Galactic halo(s) emphasizes the complexity of disrupting satellites
down to the smallest, star cluster scales. 
We note, however, that small satellites, which make up the outer halo should hardly have contributed any GCs, whereas
massive satellites contributing to the inner halo are likely to come with more GCs. 
Larger and deeper
spectroscopic samples are needed in order to verify the potential rise
seen in the outer halo in Fig.~11.

Overall, the CN-strong stars we detected in our sample are fairly
similar in many regards to their CN-weak halo mates. The latter
display a mild radial metallicity gradient, as is also seen in the GC-donated component, 
although their small numbers frustrate a unique comparison. An
important question is whether also the kinematics of these
second-generation stars coincide with those of the underlying halo
component, as this would provide valuable information on their
accretion history and the possible orbits of the progenitor systems
\citep[e.g.,][]{Roederer2018}. This will be explored in our second
paper in this series (Hanke et al., in prep.), exploiting the
capabilities of the new Gaia data \citep{GaiaDR2}.

As we have argued that the CN-strong stars broadly reflect the
dichotomy of an inner and outer halo, it can also be envisioned that
the stars at large Galactocentric distances could originate from
regular mass loss from GCs that have been accreted from satellite
dwarf galaxies \citep[cf.\ ][]{Kruijssen2015,Helmi2018GaiaEnceladus}. This tagging is enabled
by the fact that, chemically, the GC-stars in dwarf galaxies closely
follow the field star population of those dwarfs
\citep{Hendricks2016}.  In this case, further chemical tagging of the
CN-strong objects as purported (accreted) dwarf galaxy GC stars to
explore the full chemical abundance space is clearly warranted.  

Large spectroscopic surveys are bound to find ever more former GC
stars out to large distances, for instance SDSS-V
\citep{Kollmeier2017}, or 4MOST \citep{deJong2012,Helmi2019}, as their
wavelength ranges are able to capture a large number of relevant
element abundance tracers \citep{CJHansen2015}.  This goes in line
with the desirability of a broader set of abundance tracers that can
ascertain a clear-cut classification as second-generation stars.
Similarly, any observed radial trends in the significance of GC
contributions to the build-up of the inner and outer Galactic halos
will greatly benefit from the recent and future data releases of Gaia,
by pinning down the proper motions and the distances to the CN-strong
candidates, although Gaia's reach is limited towards smaller 
distances in the outer halo. 
\begin{acknowledgements}
This work was supported by Sonderforschungsbereich SFB 881 "The MW
System" (subproject A08) of the German Research Foundation (DFG).  We 
are grateful to { the referee, N. Bastian, for a very constructive report and helpful discussions.} 
SLM acknowledges support from the Australian Research Council through
Discovery Project grant DP180101791. Parts of this research were
supported by the Australian Research Council Centre of Excellence for
All-Sky Physics in 3 Dimensions (ASTRO 3D), through project number
CE170100013.
This publication has benefited from the conference on ``Multiple
populations in stellar clusters'' held at the Sexten Center for
Astrophysics (\url{http://www.sexten-cfa.eu/}).
Funding for the Sloan Digital Sky Survey IV has been provided by the
Alfred P. Sloan Foundation, the U.S. Department of Energy Office of
Science, and the Participating Institutions. SDSS acknowledges support
and resources from the Center for High-Performance Computing at the
University of Utah. The SDSS web site is www.sdss.org.
SDSS is managed by the Astrophysical Research Consortium for the
Participating Institutions of the SDSS Collaboration including the
Brazilian Participation Group, the Carnegie Institution for Science,
Carnegie Mellon University, the Chilean Participation Group, the
French Participation Group, Harvard-Smithsonian Center for
Astrophysics, Instituto de Astrof\'{\i}sica de Canarias, The Johns
Hopkins University, Kavli Institute for the Physics and Mathematics of
the Universe (IPMU) / University of Tokyo, Lawrence Berkeley National
Laboratory, Leibniz Institut f\"ur Astrophysik Potsdam (AIP),
Max-Planck-Institut f\"ur Astronomie (MPIA Heidelberg),
Max-Planck-Institut f\"ur Astrophysik (MPA Garching),
Max-Planck-Institut f\"ur Extraterrestrische Physik (MPE), National
Astronomical Observatories of China, New Mexico State University, New
York University, University of Notre Dame, Observat\'orio Nacional /
MCTI, The Ohio State University, Pennsylvania State University,
Shanghai Astronomical Observatory, United Kingdom Participation Group,
Universidad Nacional Aut\'onoma de M\'exico, University of Arizona,
University of Colorado Boulder, University of Oxford, University of
Portsmouth, University of Utah, University of Virginia, University of
Washington, University of Wisconsin, Vanderbilt University, and Yale
University.
\end{acknowledgements}
\bibliographystyle{aa} % style aa.bst
\bibliography{ms} % your references Yourfile.bib
\end{document}